\documentclass[11pt]{article}
\usepackage[margin=1in]{geometry}
\usepackage{url}
\usepackage{numprint}
\usepackage{setspace}                
\usepackage{graphicx}
\usepackage[numbers]{natbib}
\usepackage{appendix}
\usepackage{amssymb,amsmath,bbm,float}
\usepackage{epstopdf}
\usepackage{xcolor}
\usepackage{enumerate}

\usepackage{xr}

\DeclareMathOperator*{\argmax}{arg\,max}
\DeclareMathOperator*{\argmin}{arg\,min}

\newtheorem{thm}{Theorem}[section]

\begin{document}

\title{Regression modelling of interval censored data based on the adaptive ridge procedure}


\author{Olivier Bouaziz$^{1}$, Eva Lauridsen$^{2}$ and Gr\'egory Nuel$^{3}$}
\date{$^1$Laboratory MAP5, University Paris Descartes and CNRS, Sorbonne Paris Cit\'e, Paris, France \\%
	$^2$Ressource Center for Rare Oral Diseases, Copenhagen University Hospital, Rigshospitalet, Denmark\\
    $^3$LPSM, CNRS 7599, 4 place Jussieu, Paris, France\\}

\maketitle


\begin{abstract}
{\color{red}A new method for the analysis of time to ankylosis complication on a dataset of replanted teeth is proposed. In this context of left-censored, interval-censored and right-censored data, a Cox model with piecewise constant baseline hazard is introduced.} Estimation is carried out with the EM algorithm by treating the true event times as unobserved variables. This estimation procedure is shown to produce a block diagonal Hessian matrix of the baseline parameters. Taking advantage of this interesting feature of the estimation method a {\color{red}$L_0$} penalised likelihood method is implemented in order to automatically determine the number and locations of the cuts of the baseline hazard. {\color{red}This procedure allows to detect specific areas of time where patients are at greater risks for ankylosis.} The method can be directly extended to the inclusion of exact observations and to a cure fraction. Theoretical results are obtained which allow to derive statistical inference of the model parameters from asymptotic likelihood theory. Through simulation studies, the penalisation technique is shown to provide a good fit of the baseline hazard and precise estimations of the resulting regression parameters.

\noindent \textbf{Keywords}: Adaptive Ridge procedure; Cure model; EM algorithm; Interval censoring; Penalised likelihood; Piecewise constant hazard.
\end{abstract}


\section{Introduction}

Interval censored data arise in situations where the event of interest is only known to have occurred between two observation times. These types of data are commonly encountered when the patients are intermittently followed up at medical examinations. This is the case for instance in AIDS studies, when HIV infection onset is determined by periodic testing, or in oncology where the time-to-tumour progression is assessed by measuring the tumour size at periodic testing. Dental data are another examples which are usually interval-censored because the teeth status of the patients are only examined at visits to the dentist. While interval-censored data are ubiquitous in medical applications it is still a common practice to replace the observation times with their midpoints or endpoints and to consider these data as exact. This allows to analyse the data using standard survival approach but may result in a large bias of the estimators.  
{\color{red}In the present paper we develop a new method for the analysis of time to ankylosis complication on a dataset of replanted teeth. The three main goals for our method is to adequately take into account interval-censoring, to be able to identify time ranges where patients are particularly at high risk of developing the complication and to investigate if a sub-population of non susceptible patients exists.}

In the context of interval-censored data, \cite{turnbull1976} introduced an iterative algorithm for the non-parametric estimation of the survival function. As a different estimation method, the iterative convex minorant was proposed by \cite{groeneboom92} and \cite{jongbloed1998iterative}. In~\cite{groeneboom92}, the authors derived the slow rate of convergence of order $n^{1/3}$ for the non-parametric survival estimator. Moreover, the obtained law is not Gaussian and cannot be explicitly computed. Many methods were also developed in a regression setting. In particular, the Cox model with non-parametric baseline was studied in~\cite{huang1995efficient}. The authors derived a $n^{1/2}$ convergence rate for the regression parameter with a Gaussian limit but the problem of estimation and inference of the baseline survival function pertains in this regression context: the baseline survival function has the $n^{1/3}$ slow rate of convergence and even more problematic, the asymptotic distribution of this function could not be derived. The same conclusions were observed in~\cite{boruvka2015cox} where the authors use the more general Cox-Aalen model with non-parametric baseline.  As a consequence, alternatives to the non-parametric baseline have been introduced. In~\cite{lindsey1998study} and~\cite{sun07} parametric baselines such as Weibull or piecewise constant are introduced. In that case, the convergence rate of the global parameters is of order $n^{1/2}$ and the asymptotic distribution is Gaussian (see~\cite{sun07}).  In~\cite{betensky2002local} a local likelihood is implemented which results in a smooth estimation of the baseline hazard using a kernel function. However, asymptotic properties of the estimators were not derived in their work and the performance of the estimators depends on the choice of the kernel bandwidth. In~\cite{wang2016flexible}, monotone B-splines are implemented in order to estimate the cumulative baseline hazard. The authors introduce a two stage data augmentation which allows them to use the Expectation Maximisation algorithm~\citep[EM, see][]{dempster1977maximum} in order to perform estimation. Asymptotics with $n^{1/2}$ rate of convergence of the estimators are derived. However, the number and location of the splines knots are pre-determined by the user and the estimators performance depend on the choice of these tuning parameters. {\color{red} A similar two stage data augmentation approach was developed in~\cite{zeng2016maximum} where the authors study the more general class of semi-parametric transformation models, using a non-parametric baseline and allowing for time dependent covariates. The $n^{1/2}$ rate of convergence of the regression parameter is derived but the asymptotic distribution of the non-parametric baseline was not obtained.}

In this work, we study the Cox model with piecewise constant baseline hazard. Treating the unobserved true event times as missing variables we use the EM algorithm to perform estimation. As a result, the Hessian of the log-likelihood to be maximised is seen to be diagonal. This is a remarkable feature of the method that easily allows to perform estimation with the piecewise constant baseline using arbitrarily large set of cuts. In contrast, this model had been already introduced in~\cite{carstensen1996regression} and~\cite{lindsey1998study} but maximisation of the model parameters was achieved using the observed likelihood which resulted in a full rank Hessian matrix. In~\cite{carstensen1996regression} for example, the authors warn against computational issues which may force the user to reduce the number of cuts by combining adjacent intervals. Using the EM algorithm to perform estimation in the piecewise constant hazard model is new to our knowledge and easy to implement. Also, all the quantities involved in the E-step can be explicitly computed in our method, contrary to previous works (see~\cite{betensky2002local} for example) which require to approximate integrals. In comparison with~\cite{wang2016flexible} the E-step is more natural and directly applicable using the complete likelihood. Moreover, taking advantage of the sparse structure of the Hessian matrix, our method can be combined with a {\color{red}$L_0$} penalty designed to detect the location and number of cuts. This is performed through the adaptive ridge procedure, a regularisation method that was introduced in~\cite{rippe2012visualization}, \cite{NuelFrommlet} and then applied in a survival context (without covariates) in~\cite{bouaziz17}. This penalisation technique results in a flexible method where the cuts and locations of the piecewise constant baseline are automatically chosen from the data, thus providing a good compromise between purely non-parametric and parametric baseline functions. This is in contrast with existing techniques such as in~\cite{wang2016flexible} where the location and number of knots of splines basis are fixed by the user. Finally we also emphasise the advantage of the {\color{red}$L_0$} method in terms of interpretability: by detecting the relevant set of cuts of the baseline the method highlights the different regions of time where the risk of failure varies. This {\color{red}is of great interest for the dental application in order for the dentists to precisely detect time intervals where patients are at a higher risk of ankylosis.}

Another advantage of using the EM algorithm is to provide direct extensions of the Cox model. In this work we also consider the inclusion of exact data in the estimation method. This mixed case of exact and interval-censored data is usually not easy to analyse as standard methods for interval-censoring do not directly extend to exact data. However, using our method, inclusion of exact data is straightforward through the E-step and the likelihood can be decomposed into the contribution of exact and interval-censored observations. Another extension that is developed in this work is the inclusion of a fraction of non-susceptible patients. This situation is modelled using the cure model of~\cite{sy00} and~\cite{peng2000nonparametric}, with a logit link for the probability of being cured. Little attention has been paid to this model in the case of interval-censored data. In~\cite{hu2016partially} the authors consider a partially linear transformation model where the baseline is modelled using spline basis but the number and location of knots are chosen in an ad-hoc manner. In~\cite{liu2009semiparametric} a different cure model was introduced where the marginal survival function (without conditioning on the susceptible group) is modelled. However, the asymptotic distribution of the estimated parameters were not derived under this model.  With our method, estimation in the cure Cox model is straightforward. The E-step results in a weighted log-likelihood with the weights corresponding to the probability of being cured such that our estimation method readily extends to the cure model. {\color{red}This model is especially useful on the dental dataset to assess if there exists a subpopulation of patients who are not at risk of developing the ankylosis complication.}

In Section~\ref{sec:model} the piecewise constant hazard model is introduced. The estimation method based on the EM algorithm is presented in Section~\ref{sec:fixedcuts} for interval censored data and fixed cuts of the hazard. Estimation in the non-parametric case, in the regression model and  extensions for exact data and the cure model are also developed in this section. Then, the {\color{red}$L_0$} penalised likelihood that allows to select the location and number of cuts from the data is presented in Section~\ref{sec:ar}. Asymptotic properties of the penalised estimator are discussed in Section~\ref{sec:test_CI}. In particular, these results show that confidence intervals and tests can be constructed by considering the selected cuts as fixed. In Section~\ref{sec:simu}, an extensive simulation study is presented where our adaptive ridge estimator is compared with the midpoint estimator and the ICsurv estimator from~\cite{wang2016flexible}. Finally, {\color{red} the} dental dataset on {\color{red} ankylosis} complications for replanted teeth is analysed in Section~\ref{sec:data} {\color{red}using the proposed methodology}.

\section{A piecewise constant hazard model for interval censored data}\label{sec:model}

Let $T$ denote the time to occurrence of the event of interest. We consider a situation where all individuals are subject to interval censoring defined by the random variables $(L,R)$ such that $L$ and $R$ are observed and $\mathbb P(T\in[L,R])=1$. The situation $L=0$ and $R<\infty$ corresponds to left-censoring, $0<L<R<\infty$ corresponds to strictly interval censoring and $L<R=\infty$ to right censoring. The special case $L=R$ is also allowed which corresponds to exact observations of the time of interest. We introduce a {\color{red}column} covariate vector $Z$ of dimension $d_Z$ and for convenience we also introduce $\delta$ which equals $0$ if an individual is right censored and $1$ if he/she is left, interval censored {\color{red}or exactly observed}. The variable $T$ is considered continuous and we assume independent censoring in the following way (see for instance~\cite{zhang2005regression}): 
$\mathbb P(T\leq t\mid L=l,R=r,Z)=\mathbb P(T\leq t\mid l\leq T\leq r,Z).$ 
This supposes that the variables $(L,R)$ do not convey additional information on the law of $T$ apart from assuming $T$ to be bracketed by $L$ and $R$. Finally, we assume non-informative censoring in the sense that the distribution of $L$ and $R$ does not depend on the model parameters involved in the distribution of $T$.



%
%
%
We consider the following Cox proportional hazard model for the time variable $T$: 
\linespread{0.9}
\begin{align}\label{eq:Cox}
\lambda(t\mid Z)=\lambda_0(t)\exp(\beta Z),
\end{align}
where $\beta$ is an unknown row parameter vector of dimension $d_Z$. 
We model the baseline function $\lambda_0$ through a piecewise constant hazard. Let $c_0,c_1,\ldots, c_K$ represent $K+1$ cuts, with the convention that $c_0=0$ and $c_K=+\infty$. Let $I_k(t)=I(c_{k-1}<t\leq c_k)$, with $I(\cdot)$ denoting the indicator function. We suppose that
$\lambda_0(t)=\sum_{k=1}^K I_k(t)\exp(a_k).$ 
Under this model, note that the survival and density functions are respectively equal to:
\begin{align*}
S(t\mid Z)&=\exp\Big(-\sum_{k=1}^K e^{a_k+\beta Z}(t\wedge c_k-c_{k-1})I(c_{k-1}\leq t)\Big),\\
f(t\mid Z)&=\sum_{k=1}^K I_k(t)\exp\Big(a_k+\beta Z-\sum_{j=1}^k e^{a_j+\beta Z}(t\wedge c_j-c_{j-1})\Big).
\end{align*}
We set $\boldsymbol{\theta}=(a_1, \ldots, a_K,\beta)$ the model parameter we aim to estimate. In the following, we will also study the so-called nonparametric situation, when no covariates are available, which is encompassed in our modelling approach as the special case where $Z=0$. In this context the hazard function is simply equal to $\lambda_0$ which is assumed to be piecewise constant and the model parameter is $\boldsymbol{\theta}=(a_1, \ldots, a_K)$. The observed data consist of $\text{data}={\color{red}\{\text{data}_i,i=1,\ldots,n\}}$ with ${\color{red}\text{data}_i}=(L_i,R_i,\delta_i)$ in the nonparametric context and ${\color{red}\text{data}_i}=(L_i,R_i,\delta_i,Z_i)$ in the regression context, while $T_i$ is considered as incompletely observed. In the latter context, we introduce the notation $a_{i,k}=a_k+\beta Z_i$. 
  

\section{Estimation procedure with fixed cuts}\label{sec:fixedcuts}

For the sake of simplicity, we first consider the scenario when no exact data are observed (which means there only are left, interval and right censored data). The estimation method is based on the EM algorithm and is presented in Section~\ref{sec:EM} in the general regression context since the nonparametric context can be easily derived by setting $Z=0$. The nonparametric context is discussed in Section~\ref{sec:np}, the implementation of the M step for the regression context is presented in Section~\ref{sec:reg} and the method when exact observations are also available is developed in Section~\ref{sec:exact}. Finally, the inclusion of a fraction of non-susceptible individuals is studied in Section~\ref{sec:cure}.
\subsection{The EM algorithm for left, right and interval censored observations}\label{sec:EM}

The observed likelihood is defined with respect to the observed data by:
\begin{align*}
\mathrm{L}_n^{\text{obs}}(\boldsymbol\theta)&={\color{red}\prod_{i =1}^n (S(L_i\mid  Z_i,\boldsymbol\theta)-S(R_i\mid  Z_i,\boldsymbol\theta))}\\
&=\prod_{i=1}^n \left\{\exp\Big(-\int_0^{L_i} \lambda_0(t)dt \,e^{\beta Z_i}\Big)\left(1-\exp\Big(-\int_{L_i}^{R_i} \lambda_0(t)dt\, e^{\beta Z_i}\Big)\right)\right\}^{\delta_i}\\
& \qquad \times \left\{\exp\Big(-\int_0^{L_i} \lambda_0(t)dt\, e^{\beta Z_i}\Big)\right\}^{1-\delta_i},
\end{align*}
{\color{red}with the slight abuse of notation $S(R_i\mid  Z_i,\boldsymbol\theta)=0$ if $R_i=\infty$ (for a right-censored observation).}
The Maximum Likelihood Estimator (MLE) can be derived from maximisation of this observed log-likelihood with respect to the model parameters, as in~\cite{carstensen1996regression} for instance. The obtained parameter estimates are not explicit but a Newton-Raphson algorithm can be easily implemented. However, in this optimisation problem, the block of the Hessian matrix corresponding of the baseline coefficients $a_1,\ldots,a_K$ will be of full rank and can lead to intractable solutions if the number of cuts $K$ is large. An alternative method to compute the MLE is therefore to use the EM algorithm based on the complete likelihood of the unobserved true event times. This algorithm will result into a diagonal block matrix of the baseline coefficients.


The EM algorithm is based on the complete likelihood, defined by:  $\mathrm{L}_n(\boldsymbol\theta)=\prod_{i=1}^n f(T_i\mid Z_i,\boldsymbol\theta).$ 
Denote by $\boldsymbol\theta_{\text{old}}$ the current parameter value. The E-step takes the expectation of the complete log-likelihood with respect to the $T_i$'s, given the $L_i$'s, $R_i$'s, $\delta_i$'s, $Z_i$'s and $\boldsymbol\theta_{\text{old}}$. 
Write
\begin{align*}
Q_i(\boldsymbol\theta\mid \boldsymbol\theta_{\text{old}}):=\mathbb E[\log(f(T_i\mid Z_i,\boldsymbol\theta))\mid\text{data}_{{\color{red}i}},\boldsymbol\theta_{\text{old}}] & =\int f(t\mid\text{data}_{{\color{red}i}},\boldsymbol\theta_{\text{old}})\log f(t\mid Z_i,\boldsymbol\theta)dt,
\end{align*}
where $f(t\mid\text{data}_{{\color{red}i}},\boldsymbol\theta_{\text{old}})$ represents the conditional density of $T_i$ given $\text{data}_{{\color{red}i}}$ and $\boldsymbol\theta_{\text{old}}$, evaluated at $t$. Under the independent censoring assumption,
\begin{align*}
f(t\mid\text{data}_{{\color{red}i}},\boldsymbol\theta_{\text{old}})=\frac{f(t\mid Z_i,\boldsymbol\theta_{\text{old}})I(L_i<t<R_i)}{S(L_i\mid  Z_i,\boldsymbol\theta_{\text{old}})-S(R_i\mid Z_i,\boldsymbol\theta_{\text{old}})}\cdot
\end{align*}
The E-step consists of computing the quantity $Q(\boldsymbol\theta\mid \boldsymbol\theta_{\text{old}}) =\sum_i Q_i(\boldsymbol\theta\mid \boldsymbol\theta_{\text{old}})$. We have: 
\begin{align*}
Q(\boldsymbol\theta\mid \boldsymbol\theta_{\text{old}})\! &=\!\!\sum_{i=1}^n\frac{\int_{L_i}^{R_i} f(t\mid Z_i,\boldsymbol\theta_{\text{old}})\log f(t\mid Z_i;\boldsymbol\theta)dt}{{S(L_i\mid Z_i,\boldsymbol\theta_{\text{old}})-S(R_i\mid Z_i,\boldsymbol\theta_{\text{old}})}}
\end{align*}
\begin{align*}
Q(\boldsymbol\theta\mid \boldsymbol\theta_{\text{old}})\!&=\!\! \sum_{i=1}^n\Bigg\{\frac{1}{S(L_i\mid Z_i,\boldsymbol\theta_{\text{old}})-S(R_i\mid Z_i,\boldsymbol\theta_{\text{old}})}\\
\!&\!\!\quad\!\!\!\!\!\times \sum_{k=1}^K J_{k,i}\!\!\int_{c_{k-1}\vee L_i}^{c_k\wedge R_i}\!\! \exp\Big(a_{i,k}^{{\text{old}}}-\sum_{j=1}^{k}e^{a_{i,j}^{\text{old}}}(t\wedge c_j-c_{j-1})\Big)\!\Big(a_{i,k}-\sum_{j=1}^{k} e^{a_{j,k}} (t\wedge c_j-c_{j-1})\Big)dt\!\Bigg\},
\end{align*}
where $J_{k,i}$ is the indicator $I\{(L_i,R_i)\cap(c_{k-1},c_k)\neq \emptyset\}$ and $b_1\wedge b_2$, $b_1 \vee b_2$ respectively denote $\min(b_1,b_2)$, $\max(b_1,b_2)$. 
Finally, the M-step corresponds of maximising, with respect to $\boldsymbol\theta$, the quantity
\begin{align*}
Q(\boldsymbol\theta\mid \boldsymbol\theta_{\text{old}})= \sum_{i=1}^n\sum_{k=1}^K\bigg\{\Big(a_{i,k}-\sum_{j=1}^{k-1}(c_j-c_{j-1})e^{a_{i,j}}\Big) A^{\text{old}}_{k,i}-e^{a_{i,k}}B^{\text{old}}_{k,i}\bigg\},
\end{align*}
where exact expressions of the statistics $A^{\text{old}}_{k,i}$ and $B^{\text{old}}_{k,i}$ can be found in the Supplementary Material.\\


\subsection{Estimation in the absence of covariates}\label{sec:np}

In the absence of covariates, the previous results hold with $Z_i=0$, $a_{i,k}=a_k$ and the model parameters we aim to estimate are just $\boldsymbol{\theta}=(a_1, \ldots, a_K)$. The objective function in the M-step can be defined with respect to 
the sufficient statistics $\bar A^{\text{old}}_{k}= \sum_i A^{\text{old}}_{k,i}$ and $\bar B^{\text{old}}_{k}=\sum_i B^{\text{old}}_{k,i}$:
 \begin{align*}
Q(\boldsymbol\theta\mid \boldsymbol\theta_{\text{old}})= \sum_{k=1}^K\bigg\{\Big(a_{k}-\sum_{j=1}^{k-1}(c_j-c_{j-1})e^{a_{j}}\Big) \bar A^{\text{old}}_{k}-e^{a_{k}}\bar B^{\text{old}}_{k}\bigg\}.
\end{align*}
The derivatives of $Q$ with respect to $a_k$, $k=1,\ldots,K$, equal
\begin{align*}
\frac{\partial Q(\boldsymbol\theta\mid \boldsymbol\theta_{\text{old}})}{\partial a_k} & = \bar A^{\text{old}}_{k}-(c_k-c_{k-1})e^{a_k}I(k\neq K)\sum_{l=k+1}^K \bar A^{\text{old}}_{l}-e^{a_k}\bar B^{\text{old}}_{k}.
\end{align*}
As a consequence, in the absence of covariates, one gets the explicit parameters estimators:
\begin{align*}
\exp(\hat a_k) & = \frac{\bar A^{\text{old}}_{k}}{I(k\neq K)\sum_{l=k+1}^K \bar A^{\text{old}}_{l}(c_k-c_{k-1})+\bar B^{\text{old}}_{k}}, k=1,\ldots,K,
\end{align*}
at each step of the EM algorithm. At convergence, this provides an estimator of the hazard function from which quantities of interest, such as the survival function, can be easily derived.

\subsection{Estimation in the general regression framework}\label{sec:reg}

In the regression framework, each step of the EM algorithm is solved through a Newton-Raphson procedure. 
The first and second order derivatives of $Q$ with respect to $a_k$ and $\beta$ are equal to 
\begin{align*}
\frac{\partial Q(\boldsymbol\theta\mid \boldsymbol\theta_{\text{old}})}{\partial a_k} & = \sum_{i=1}^n\left\{A^{\text{old}}_{k,i}-(c_k-c_{k-1})e^{a_{k}}I(k\neq K)\sum_{l=k+1}^KA^{\text{old}}_{l,i}e^{\beta Z_i}-e^{a_{k}}B^{\text{old}}_{k,i}e^{\beta Z_i}\right\},\\
\frac{\partial Q(\boldsymbol\theta\mid \boldsymbol\theta_{\text{old}})}{\partial \beta} & = \sum_{i=1}^nZ_i\sum_{l=1}^K\left(A^{\text{old}}_{l,i}-\Bigg\{\sum_{j=1}^{l-1}(c_j-c_{j-1})e^{a_{j}}A^{\text{old}}_{l,i}e^{\beta Z_i}+e^{a_{l}}B^{\text{old}}_{l,i}e^{\beta Z_i}\Bigg\}\right),
\end{align*}
and
\begin{align*}
\frac{\partial^2 Q(\boldsymbol\theta\mid \boldsymbol\theta_{\text{old}})}{\partial a^2_k} & = -\sum_{i=1}^n\left\{(c_k-c_{k-1})e^{a_{k}}I(k\neq K)\sum_{l=k+1}^K A^{\text{old}}_{l,i}e^{\beta Z_i}+e^{a_{k}}B^{\text{old}}_{k,i}e^{\beta Z_i}\right\},\\
\frac{\partial^2 Q(\boldsymbol\theta\mid \boldsymbol\theta_{\text{old}})}{\partial \beta^2} & = -\sum_{i=1}^nZ_iZ_i^t\sum_{l=1}^K\left(\sum_{j=1}^{l-1}(c_j-c_{j-1})e^{a_{j}}A^{\text{old}}_{l,i}e^{\beta Z_i}+e^{a_{l}}B^{\text{old}}_{l,i}e^{\beta Z_i}\right),\\
\frac{\partial^2 Q(\boldsymbol\theta\mid \boldsymbol\theta_{\text{old}})}{\partial a_k\partial \beta} & = -\sum_{i=1}^nZ_i\left((c_k-c_{k-1})e^{a_{k}}I(k\neq K)\sum_{l=k+1}^{K}A^{\text{old}}_{l,i}e^{\beta Z_i}+e^{a_{k}}B^{\text{old}}_{k,i}e^{\beta Z_i}\right).
\end{align*}
The block matrix of the Hessian corresponding to the second order derivatives with respect to the $a_k$'s is diagonal while the three other blocks are of full rank. Inversion of the Hessian matrix is then achieved using the Schurr complement which takes advantage of this sparse structure of the Hessian. When considering a large number of cuts, that is $K>>d_Z$, the total complexity of the inversion of the Hessian is of order $\mathcal{O}(K)$. The exact formula of the Schurr complement is given in the Supplementary Material.

\subsection{Inclusion of exact observations}\label{sec:exact}

It is straightforward to deal with exact observations since they can be directly included in the EM algorithm. For an exact observation $i$, $\mathbb E[\log(f(T_i\mid Z_i;\boldsymbol\theta))\mid \text{data},\boldsymbol\theta_{\text{old}}] =\log(f(T_i\mid Z_i;\boldsymbol\theta))=\sum_{k=1}^K \big\{O_{i,k}a_{i,k}-\exp(a_{i,k})R_{i,k}\big\},$ 
with $O_{i,k}=I(c_{k-1}<T_i<c_k)$ and $R_{i,k}=T_i\wedge c_k-c_{k-1}$. Note that this corresponds to the classical contribution of an exact observation to the log-likelihood in the standard Poisson regression for right censored observations (see for instance~\cite{aalen_borgan_book}). As a result, $Q$ can be decomposed as
\begin{align*}
Q(\boldsymbol\theta\mid \boldsymbol\theta_{\text{old}}) & = \sum_{i \text{ not exact}}\sum_{k=1}^K\bigg\{\Big(a_{i,k}-\sum_{j=1}^{k-1}(c_j-c_{j-1})e^{a_{i,j}}\Big) A^{\text{old}}_{k,i}-e^{a_{i,k}}B^{\text{old}}_{k,i}\bigg\}\\
& \quad +\sum_{i \text{ exact}}\sum_{k=1}^K\bigg\{O_{i,k}a_{i,k}-\exp(a_{i,k})R_{i,k}\bigg\}.
\end{align*}
The estimation method follows as previously. In particular, in the absence of covariates, the explicit parameters estimator of $(a_1, \ldots, a_K)$ are equal to:
\begin{align*}
\exp(\hat a_k) & = \frac{\bar A^{\text{old}}_{k}+\bar O_{k}}{I(k\neq K)\sum_{l=k+1}^K \bar A^{\text{old}}_{l}(c_k-c_{k-1})+\bar B^{\text{old}}_{k}+\bar R_{k}}, k=1,\ldots,K,
\end{align*}
where $\bar O_{k}=\sum_{i \text{ exact}}\bar O_{i,k}$ and $\bar R_{k}=\sum_{i \text{ exact}}\bar R_{i,k}$.

In the regression setting, maximisation over the $\beta$ and $a_1,\ldots,a_K$ parameters is performed through the Newton-Raphson algorithm as before. Full expressions of the score vector and Hessian matrix are given in the Supplementary Material. The Schurr complement is used again to invert the Hessian matrix (see the Supplementary Material).


\subsection{Inclusion of a fraction of non-susceptibles (cure fraction)}\label{sec:cure}

Taking into account non-susceptible individuals is possible using the cure model from~\cite{sy00}. This is achieved by modelling the latent status (susceptible/non-susceptible) of the individuals through a variable $Y$ which equals $1$ for patients that will eventually experience the event and $0$ for patients that will never experience the event. Since the estimation method uses the EM algorithm, this latent variable can be easily dealt with through the E-step.

We assume that $Y$ is independent of $T$ conditionally on $(L,R)$. 
The proportional hazard Cox model for the susceptibles is defined as 
\begin{align}\label{eq:CoxCure}
\lambda(t\mid Y=1,Z)=\lambda_0(t)\exp(\beta Z).
\end{align}
The cure model specifies the hazard, conditional on $Y$ and $Z$, to be equal to $\lambda(t\mid Y,Z)=Y\lambda(t\mid Y=1,Z)$. The baseline function $\lambda_0$ is assumed to be piecewise constant as in Section~\ref{sec:model} and the conditional density and survival functions of the susceptibles are respectively noted $f(t\mid Y=1,Z)$ and $S(t\mid Y=1,Z)$. If one wants to model the effect of covariates on the probability of being cured, a logistic link can be used:
\begin{align}\label{eq:probcure}
p(X)= \mathbb P[Y = 1\mid X]=\frac{\exp(\gamma X)}{1+\exp(\gamma X)},
\end{align}
where $X$ is a covariate vector including the intercept and $\gamma$ is a row parameter vector, both of dimension $d_X$. The observed data then consist of $\text{data}=(L_i,R_i,\delta_i,Z_i,X_i)_{1\leq i\leq n}$ while $T_i$ and $Y_i$ are respectively incompletely observed and non observed data. The model parameter is $\boldsymbol{\theta}=(a_1, \ldots, a_L,p)$ in the completely nonparametric context (no covariates $X$ nor $Z$), $\boldsymbol{\theta}=(a_1, \ldots, a_L,\beta,p)$ if only the covariate $Z$ is used or $\boldsymbol{\theta}=(a_1, \ldots, a_L,\beta,\gamma)$ in the full regression context (with covariates $X$ and $Z$). In the later case, we introduce the notation $p_i=\mathbb P[Y_i = 1\mid X_i]$. The other situations are encompassed in our modelling approach by setting $X=0$ and/or $Z=0$. Note that our cure model is identifiable and does not require additional constraints such as in~\cite{sy00} where the authors had to impose $S(t\mid Y=1,Z)$ to be null for $t$ greater than the last event time in the context of exact and right-censored data.

Under the cure model with interval-censored and exact observations, the observed likelihood is now defined as
\begin{align*}
 \mathrm{L}_n^{\text{obs}}(\boldsymbol\theta)&=\prod_{i \text{ not exact}} \left\{p_i\exp\Big(-\int_0^{L_i} \lambda_0(t)dt e^{\beta_0 Z_i}\Big)\left(1-\exp\Big(-\int_{L_i}^{R_i} \lambda_0(t)dt e^{\beta_0 Z_i}\Big)\right)\right\}^{\delta_i}\\
& \qquad \times \left\{(1-p_i)+p_i\exp\Big(-\int_0^{L_i} \lambda_0(t)dt e^{\beta_0 Z_i}\Big)\right\}^{1-\delta_i}\prod_{i \text{ exact}} p_i f(T_i\mid Y_i=1,Z_i;\boldsymbol\theta)
\end{align*}
and the complete likelihood is defined as: $\mathrm{L}_n(\boldsymbol\theta)=\prod_{i=1}^n p_i^{Y_i}(1-p_i)^{1-Y_i}\{f(T_i\mid Y_i=1,Z_i;\boldsymbol\theta)\}^{Y_i}.$ 
The E-step consists of computing the function $Q(\boldsymbol\theta\mid \boldsymbol\theta_{\text{old}}) =\mathbb E[\log(\mathrm{L}_n(\boldsymbol\theta))\mid  \text{data},\boldsymbol\theta_{\text{old}}]$. Let $\pi_i^{\text{old}}=\mathbb E[Y_i\mid \text{data},\boldsymbol\theta_{\text{old}}]$, we have:
\begin{align*}
\pi_i^{\text{old}}=\delta_i+\frac{(1-\delta_i)p_{\text{old}}S(L_i\mid Y_i=1,Z_i,\boldsymbol\theta_{\text{old}})}{1-p_{\text{old}}+p_{\text{old}}S(L_i\mid Y_i=1,Z_i,\boldsymbol\theta_{\text{old}})}\cdot
\end{align*}
In the case of interval-censored and exact observations, 
\begin{align*}
Q(\boldsymbol\theta\mid \boldsymbol\theta_{\text{old}}) & = \sum_{i=1}^n\left\{\pi_i^{\text{old}}\log(p_i)+(1-\pi_i^{\text{old}})\log(1-p_i)\right\}\\
& \quad +\sum_{i \text{ not exact}}\pi_i^{\text{old}}\sum_{k=1}^K\bigg\{\Big(a_{i,k}-\sum_{j=1}^{k-1}(c_j-c_{j-1})e^{a_{i,j}}\Big) A^{\text{old}}_{k,i}-e^{a_{i,k}}B^{\text{old}}_{k,i}\bigg\}\\
& \quad +\sum_{i \text{ exact}}\sum_{k=1}^K\bigg\{O_{i,k}a_{i,k}-\exp(a_{i,k})R_{i,k}\bigg\},
\end{align*}
where $A^{\text{old}}_{k,i}$, $B^{\text{old}}_{k,i}$ are defined as in the Supplementary Material with the quantity $S(\cdot\mid Z_i,\boldsymbol\theta_{\text{old}})$ replaced by $S(\cdot\mid Y_i=1,Z_i,\boldsymbol\theta_{\text{old}})$. The terms $O_{i,k}$ and $R_{i,k}$ were defined in Section~\ref{sec:exact}.

The $Q$ function separates the terms with $\gamma$ and the terms involving $(a_1, \ldots, a_K,\beta)$ such that maximisation of these terms can be performed separately. Let $\bar A^{\pi,\text{old}}_{k}=\sum_{i}\pi_i^{\text{old}} A^{\text{old}}_{k,i}$, $\bar B^{\pi,\text{old}}_{k}=\sum_i \pi_i^{\text{old}}B^{\text{old}}_{k,i}$ and $\bar \pi^{\text{old}}=\sum_i  \pi_i^{\text{old}}$. In the nonparametric setting, explicit estimators of the parameters can be computed at each step of the EM algorithm through the formulas:
\begin{align*}
\hat p & =\frac{\bar \pi^{\text{old}}}{n},\\
\exp(\hat a_k) & = \frac{\bar A^{\pi,\text{old}}_{k}+\bar O_{k}}{I(k\neq K)\sum_{l=k+1}^K \bar A^{\pi,\text{old}}_{l}(c_k-c_{k-1})+\bar B^{\pi,\text{old}}_{k}+\bar R_{k}}, k=1,\ldots,K.
\end{align*}
In the general regression context, a Newton-Raphson procedure is implemented separately to maximise both parts of $Q$. The first and second order derivatives of $Q$ with respect to $\gamma$ are equal to:
\begin{align*}
\frac{\partial Q(\boldsymbol\theta\mid \boldsymbol\theta_{\text{old}})}{\partial \gamma} & =\sum_{i=1}^nX_i\left(\pi_i^{\text{old}}-\frac{\exp(\gamma X_i)}{1+\exp(\gamma X_i)}\right),\\
\frac{\partial^2 Q(\boldsymbol\theta\mid \boldsymbol\theta_{\text{old}})}{\partial \gamma^2} &=-\sum_{i=1}^nX_iX_i^t\frac{\exp(\gamma X_i)}{(1+\exp(\gamma X_i))^2}\cdot
\end{align*}
Exact expressions of the first and second order derivatives of $Q$ with respect to $a_k$ and $\beta$ are given in the Supplementary Material. They are expressed as weighted versions with respect to $\pi_i^{\text{old}}$ of the derivatives obtained in the context where all individuals are susceptibles. As previously, the block matrix corresponding to the second order derivatives with respect to the $a_k$s of the Hessian is diagonal and inversion of the Hessian matrix is achieved using the Schurr complement.

\section{Estimation procedure using the adaptive ridge method}\label{sec:ar}

In this section we present a penalised estimation method to detect the number and location of the cuts of the baseline hazard, when those are not known in advance. The proposed methodology is based on the work of~\cite{rippe2012visualization}, \cite{NuelFrommlet} and~\cite{bouaziz17} and can be applied to any of the previous scenarios (with exact observations, with a cure fraction, in a nonparametric setting, in a regression setting) where the function $Q$ represents the objective function associated with the context under study.

\subsection{A penalised EM algorithm}

If the number of cuts is not known in advance, we choose a large grid of cuts (i.e $K$ large) and we penalise the log-likelihood in the manner of~\cite{NuelFrommlet}, \cite{rippe2012visualization} and~\cite{bouaziz17}. This penalisation is designed to enforce consecutive values of the $a_k$s that are close to each other to be equal. It is defined in the following way:
\begin{align}\label{eq:penlikeli}
\ell^{\text{pen}}(\boldsymbol\theta\mid \boldsymbol\theta_{\text{old}})=Q(\boldsymbol\theta\mid \boldsymbol\theta_{\text{old}})-\frac{\text{pen}}{2}\sum_{k=1}^{K-1}\hat w_k(a_{k+1}-a_k)^2,
\end{align}
where $\boldsymbol{\hat w}=(\hat w_1,\ldots,\hat w_{K-1})$ are non-negative weights that will be iteratively updated in order for the weighted ridge penalty term to approximate the {\color{red}$L_0$} penalty. The pen term is a tuning parameter that describes the degree of penalisation. Note that the two extreme situations pen$=0$ and pen$=\infty$ respectively correspond to the unpenalised log-likelihood model of Section~\ref{sec:fixedcuts} and to the Cox model with exponential baseline.

Only the maximisation over $(a_1,\ldots,a_K)$ is affected by the penalty. The first and second order derivatives of $\ell^{\text{pen}}$ with respect to $a_1,\ldots,a_K$ are equal to:
\begin{align*}
\frac{\partial \ell^{\text{pen}}(\boldsymbol\theta\mid \boldsymbol\theta_{\text{old}})}{\partial a_k} & =\frac{\partial Q(\boldsymbol\theta\mid \boldsymbol\theta_{\text{old}})}{\partial a_k}
+(\hat w_{k-1}a_{k-1}-(\hat w_{k-1}+\hat w_k)a_k+\hat w_ka_{k+1})\text{pen},\\
\frac{\partial^2 \ell^{\text{pen}}(\boldsymbol\theta\mid \boldsymbol\theta_{\text{old}})}{\partial a_k^2} & = \frac{\partial^2 Q(\boldsymbol\theta\mid \boldsymbol\theta_{\text{old}})}{\partial a^2_k} 
-(\hat w_{k-1}+\hat w_k)\text{pen},
\end{align*}
\begin{align*}
\frac{\partial^2 \ell^{\text{pen}}(\boldsymbol\theta\mid \boldsymbol\theta_{\text{old}})}{\partial a_ka_{k+1}} & =\frac{\partial^2 \ell^{\text{pen}}(\boldsymbol\theta\mid \boldsymbol\theta_{\text{old}})}{\partial a_{k+1}a_k}=  
\hat w_k\text{pen},\\
\frac{\partial^2 \ell^{\text{pen}}(\boldsymbol\theta\mid \boldsymbol\theta_{\text{old}})}{\partial a_ka_{k'}} & =0 \text{ for } k, k' \text{ such that }\mid k-k'\mid \geq 2.
\end{align*}
The block matrix corresponding to the second order derivatives with respect to the $a_k$s is therefore tridiagonal. 
For a given value of pen and of the weight vector $\boldsymbol{\hat w}$, inversion of the Hessian matrix is performed using the Schurr complement as previously (see the Supplementary Material) and the Newton-Raphson algorithm is implemented to derive $\boldsymbol{\hat \theta}$. 
Once the Newton-Raphson algorithm has reached convergence, the weights are updated at the $l$th step from the equation
\begin{align}\label{eq:ARalgo}
\hat w_k^{(l)}=\left((\hat a_{k+1}^{(l)}-\hat a_k^{(l)})^2+\varepsilon^2\right)^{-1},
\end{align}
for $k=1,\ldots,K-1$ with $\varepsilon=10^{-5}$ (recommended value from~\cite{NuelFrommlet}) and where the $\hat a_k^{(l)}$'s represent the estimates of the $a_k$'s obtained through the Newton-Raphson algorithm. This form of weights is motivated by the fact that $w_k(a_{k+1}-a_k)^2$ is close to $0$ when $\mid a_{k+1}-a_k \mid <\varepsilon$ and close to $1$ when $\mid a_{k+1}-a_k \mid >\varepsilon$. Hence the penalty term tends to approximate the {\color{red}$L_0$} norm. The weights are initialized by $\hat w_k^{(0)}=1$, which gives the standard ridge estimate of $\boldsymbol{a}$. 

Finally, for a given value of pen, once the adaptive ridge algorithm has reached convergence, a set of cuts is found for the $\hat a_k$'s verifying $\hat w_k(\hat a_{k+1}-\hat a_k)^2>0.99$. {\color{red}This hard thresholding allows to provide a sparse collection of cuts}. The non-penalised log-likelihood $Q$ is then maximised using this set of cuts and the final maximum likelihood estimate is derived using the results of Section~\ref{sec:fixedcuts}. It is important to stress that the penalised likelihood is used only to select a set of cuts. Reimplementing the non-penalised log-likelihood $Q$ in the final step enables to reduce the bias classically induced by penalised maximisation techniques. 

\subsection{Choice of the penalty term}\label{sec:bic}

A Bayesian Information Criterion (BIC) is introduced in order to choose the penalty term. As explained in the previous section, for each penalty value the penalised EM likelihood~\eqref{eq:penlikeli} selects a set of cuts. For a selected set of cuts we denote by $m$ the total number of parameters to be estimated and by $\hat{\boldsymbol\theta}_{m}$ the corresponding non-penalised estimated model parameter obtained by maximisation of the $Q$ function. The BIC is then defined as:
$\text{BIC}(m)=-2 \log(\mathrm{L}_n^{\text{obs}}(\hat{\boldsymbol\theta}_{m})) + m\log (n).$

Note that the BIC is expressed here in terms of selected models. Since different penalty values can yield the same selection of cuts, the BIC needs only to be computed for all different selected models (and not for all different penalties). As an illustration of the model selection procedure, a full regularisation path is displayed in Section~\ref{sec:full_reg_path} of the Supplementary Material on a simulated data sample, where for each penalty value correspond a selection of cuts and parameter estimates. 
The final set of cuts along with its estimator $\hat{\boldsymbol\theta}_{\hat m}$ is chosen such that $\text{BIC}(\hat m)$ is minimal.



\section{Asymptotic results}\label{sec:test_CI}

Theoretical properties of the derived estimator are presented in this section for interval-censored observations which can also include exact data. Theoretical results for the cure model are omitted for the sake of presentation. Two main results are established: it is first shown that the penalised estimator asymptotically detects the true support of the baseline, in the case where the true baseline is piecewise constant and the grid used to implement the estimator contains the true cuts of the baseline hasard. In the second step of the algorithm, using the cuts obtained from the penalised estimator, the non-penalised estimator from Section~\ref{sec:fixedcuts} is implemented. It is then shown that the resulting estimator is asymptotically normal and unbiased. The limiting variance is optimal in the sense that it is equal to the variance one would obtain from implementing the non-penalised estimator with the true cuts. 

In the presence of interval-censored and exact data, the observed likelihood is equal to:
\begin{align*}
\mathrm{L}_n^{\text{obs}}(\boldsymbol\theta)&=\prod_{i \text{ not exact}} (S(L_i\mid  Z_i,\boldsymbol\theta)-S(R_i\mid  Z_i,\boldsymbol\theta))\prod_{i \text{ exact}}f(T_i\mid Z_i,\boldsymbol\theta),
\end{align*}
with the slight abuse of notation $S(R_i\mid  Z_i,\boldsymbol\theta)=0$ if $R_i=\infty$ (for a right-censored observation).
We assume that the EM procedure converges which entails that the penalised estimator that maximises Equation~\eqref{eq:penlikeli} also verifies
\begin{align}\label{eq:penobs}
\boldsymbol{\hat\theta}=(\hat a_1, \ldots, \hat a_K,\hat \beta)=\argmax_{\boldsymbol\theta\in\mathbb R^{K+d_Z}} \left\{\log(\mathrm{L}_n^{\text{obs}}(\boldsymbol\theta))-\frac{\text{pen}}{2}\sum_{k=1}^{K-1}\hat w_k^{(1)}(a_{k+1}-a_k)^2\right\}.
\end{align}
In the above formula, we consider only one iteration of the adaptive ridge procedure~\eqref{eq:ARalgo} where $\boldsymbol{\hat a}^{(1)}$ is supposed to be a consistent estimator (for example the unpenalised estimator or the ridge 
estimator). We now define a true parameter $\boldsymbol{\theta^*}=(a^*_{1}, \ldots, a^*_{K^*},\beta^*)$ which is assumed to be in a compact set and a true baseline hazard function $\lambda^*_0(t)=\sum_{k=1}^{K^*} I(c^*_{k-1}<t\leq c^*_k)\exp(a^*_k)$ with true cuts $\mathcal A^*=\{c^*_1,\ldots,c^*_{K^*}\}$. Solving~\eqref{eq:penobs} provides, after detecting the consecutive values of $\hat a_k$ that are equal, an estimated set of cuts denoted $\mathcal A_n=\{\hat c_1,\ldots,\hat c_{\hat K}\}$. Note that the size of $\mathcal A_n$ and $\mathcal A^*$ might be different and typically smaller than $K$. The unpenalised estimator obtained when using $\mathcal A_n$ is noted $\boldsymbol{\hat{\hat\theta}}_{\mathcal A_n}=(\hat{\hat a}_{1,\mathcal A_n}, \ldots, \hat{\hat a}_{\hat K,\mathcal A_n},\hat{\hat \beta}_{\mathcal A_n})$. We also define $\hat{\hat\lambda}_{0,\mathcal A_n}(t)=\sum_{k=1}^{\hat K} I(\hat c_{k-1}<t\leq \hat c_k)\exp(\hat{\hat a}_{k,\mathcal A_n})$. In order to state our theorem we first introduce 
\begin{align*}
h^*_{\boldsymbol\theta}(L_i,R_i,Z_i)=I(L_i\neq R_i) \log(S^*(L_i\mid  Z_i,\boldsymbol\theta)-S^*(R_i\mid  Z_i,\boldsymbol\theta)) + I(L_i=R_i) \log(f^*(T_i\mid Z_i,\boldsymbol\theta))
\end{align*}
and the matrices $\Sigma=-\mathbb E[\nabla_{\boldsymbol \theta}^2 h^*_{\boldsymbol\theta}(L_i,R_i,Z_i))|_{\boldsymbol\theta=\boldsymbol{\theta}^*}]$ of dimension $(K^*+d_Z)\times (K^*+d_Z)$ and $\Sigma_{\beta^*}=\{\Sigma_{i,j} : K^*+1\leq i\leq K^*+d_Z,K^*+1\leq j\leq K^*+d_Z\}$ . In the formulas, $S^*$ and $f^*$ represent the survival and density functions computed using the true set of cuts for a $\boldsymbol\theta$ of dimension $K^*+d_Z$.  
Finally we let $\tau$ represents the endpoint of the study. 



\begin{thm}\label{theo:result}
Assume that $\mathcal A^*\subset \{c_1,\ldots,c_K\}$, $\mathbb P[\{R>\tau, R<\infty\}\cup \{L>\tau\}]>0$, $Z$ is almost surely bounded and $\Sigma$ is a non-singular matrix. 
Then, if $\text{pen}/\sqrt n\to 0$ as $n\to\infty$ we have:
\begin{enumerate}
\item $\lim_{n\to\infty}\mathbb P[\mathcal A_n=\mathcal A^*]=1$.
\item for all $t\in[0,\tau]$, $\sqrt n (\hat{\hat\lambda}_{0,\mathcal A_n}(t)-\lambda^*_0(t))$ converges in distribution toward a centered Gaussian variable with variance equal to $\sum_{k=1}^{K^*} I(c^*_{k-1}<t\leq c^*_k)\exp(a^*_{k})(\Sigma_{k,k})^{-1}$.
\item $\sqrt n (\hat{\hat \beta}_{\mathcal A_n}-\beta^*)$ converges in distribution toward a centered Gaussian variable with variance equal to $(\Sigma_{\beta^*})^{-1}$.
\end{enumerate}
\end{thm}

Two important remarks can be made from this theorem. Firstly, the asymptotic variances in $2.$ and $3.$ are identical to the variances obtained in the parametric piecewise constant hazard model using the true cuts. Secondly, these two variances can be consistently estimated by
\begin{align*}
-n\times\sum_{k=1}^{\hat K} I(\hat c_{k-1}<t\leq \hat c_k)\exp(\hat{\hat a}_{k,\mathcal A_n})(\partial^2 \log(\mathrm{L}_{\mathcal A_n}^{\text{obs}}(\boldsymbol{\hat{\hat\theta}}_{\mathcal A_n}))/\partial a_k^2)^{-1},
\end{align*}
and
\begin{align*}
-n(\nabla_{\beta}^2 \log(\mathrm{L}_{\mathcal A_n}^{\text{obs}}(\boldsymbol{\hat{\hat\theta}}_{\mathcal A_n})))^{-1},
\end{align*}
where $\mathrm{L}_{\mathcal A_n}^{\text{obs}}(\boldsymbol{\hat{\hat\theta}}_{\mathcal A_n})$ represents the observed likelihood evaluated at the estimated parameter $\boldsymbol{\hat{\hat\theta}}_{\mathcal A_n}$ with the estimated cuts. In other words, this theorem states that inference on the model parameters can be achieved after selection of the cuts of the baseline function by considering these cuts as fixed parameters. The proof of the theorem is inspired from~\cite{zou2006adaptive} and is provided in the Supplementary Materials.

A direct method for deriving confidence intervals or statistical tests can therefore be based on the normal approximation of the model parameter after computing the Hessian matrix of the observed log-likelihood. 
However since the calculation of the Hessian matrix is tedious under the piecewise constant hazard model, we prefer to use a likelihood ratio test approach. This approach and the explicit expression of the Hessian are detailed in the Supplementary Material. See also~\cite{zhou2015empirical} for more details about the likelihood ratio test approach for constructing confidence intervals. Finally, note that bootstrap methods can also be implemented to derive confidence intervals. This technique is particularly relevant when the interest lies in the estimation of the survival function in a non-parametric or regression context. In order to derive the asymptotic distribution of such functional one would need to use the delta-method which may result in complicated formula for the variance estimator. The bootstrap alternative avoids these technicalities.


\section{Simulation study}\label{sec:simu}

In this section we study the performance of the proposed estimators on simulated data. In what follows, two models including two scenarios with exact, left, interval-censored and right-censored data are presented. More scenarios considering the inclusion of a cure fraction can be found in the Supplementary Material. 

We consider the Cox regression setting of Equation~\eqref{eq:Cox} where the aim is to correctly estimate the regression coefficient $\beta$ and the baseline function $\lambda_0$. We set the baseline 
as a piecewise constant function with three cuts in Model M1 and as a Weibull function in Model M2 in the following way:
\begin{align*}
\text{M1: }\,\lambda_0(t)&=\Big(0.5 \,I(0<t\leq 20)+I(20<t\leq 40)+2\, I(40<t\leq 50)+4\,I(50<t)\Big)\!\cdot \!10^{-2}\\
\text{M2: }\,\lambda_0(t)&= \frac \mu \kappa \left(\frac \mu \kappa\right)^{(\mu-1)}, \text{ }\mu=8,\, \kappa=50.
\end{align*}
In both models, the covariate vector $Z$ is of dimension $d_Z=2$ with the first component simulated as a Bernoulli variable with parameter $0.6$ and the second component is independently simulated as a uniform variable with parameters $[0,2]$. The regression parameter is equal to $\beta=(\log(2),\log(0.8))$. The values of $L_i$ and $R_i$ were determined through a visit process defined in the following way. Let $\mathcal U$ denote the uniform distribution. Two visits were simulated such that the first one $V_1\sim \mathcal U[0,60]$ and the other one $V_2=V_{1}+\mathcal U[0,120]$. Then the observations for which $T_i<V_1$ correspond to left-censored observations with $L_i=0$ and $R_i=V_1$, the observations for which $T_i>V_2$ correspond to right-censored observations with $L_i=V_2$ and $R_i=\infty$, and the observations for which $V_1<T_i<V_{2}$ correspond to strictly interval-censored observations with $L_i=V_1$ and $R_i=V_{2}$. This simulation setting corresponds to Scenario S$1$ and gave a proportion of $25\%$ of left-censored observations, $52\%$ of interval-censored observations and $23\%$ of right-censored observations in Model M$1$ and a proportion of $2\%$ of left-censored observations, $76\%$ of interval-censored observations and $22\%$ of right-censored observations in Model M$2$. In Scenario S$2$, $18\%$ of exact observations were first sampled and then the same simulation scheme for the visit process was used. The percentage of right-censored observations remains identical under this scenario for both models. 

Our adaptive ridge estimator was constructed from a grid of cuts ranging from $c_0=10$ to $c_{17}=90$, with all cuts equally spaced of size $5$. The set of penalty terms was taken, on the log scale, as the set of $200$ equally spaced values ranging from $\log(0.1)$ to $\log(10\,000)$. For the EM algorithm, the $a_k$ and $\beta$ parameters were initialised to $0$. As described in Section~\ref{sec:ar}, the BIC was used to find an estimated set of cuts and the non penalised estimator was reimplemented with this set of cuts in order to derive our final estimator. This estimator was compared with the midpoint estimator and the ICsurv estimator from~\cite{wang2016flexible}. The midpoint estimator consists of replacing the interval-censored observations by their midpoint $(L_i+R_i)/2$. The data then consist of exact and right-censored observations and can be dealt with by implementing the standard Cox regression estimators. The ICsurv estimator models the cumulative baseline function using monotone splines and uses a two-stage data augmentation method to perform estimation through the EM algorithm. {\color{red}This estimator is implemented using a more recent version of the \texttt{fast.PH.ICsurv.EM} function provided from the maintainer of the \texttt{ICsurv} package. 
Following the guidelines from {\color{red}the maintainer of the \texttt{ICsurv} package} this estimator was computed using basis splines having degree $3$ with $5$ interior knots placed evenly across the range of endpoints of the observed intervals. The $\beta$ parameters and the spline coefficients were respectively initialised to $0$ and $1$. A very fine grid of time was used for the calculation of the cumulative baseline hazard from time $0$ to time $200$ with a step equal to $0.1$. This estimator cannot include exact observations and is computed only for the Scenario S$1$ in Models M$1$ and M$2$.}


A total of $M=500$ replications were implemented and the bias and the empirical standard error (SE) of $\hat\beta$ were computed for each estimator. Confidence intervals at the $95\%$ level were constructed for $\hat\beta$ using the likelihood ratio test approach, as described in the Supplementary Material (see also Section~\ref{sec:test_CI}), and the coverage probability (CP) was reported. In order to assess the quality of estimation of $\lambda_0$, the baseline survival function $S_0(t)=\exp(-\int_0^t \lambda_0(u)du)$ was also estimated with each estimator. Then, as a measure of precision, the Integrated Mean Squared Error (MISE) was decomposed as $\textrm{MISE}(\hat S_0)=\textrm{IBias}^2(\hat S_0)+\textrm{IVar}(\hat S_0)$, where
\begin{align*}
\textrm{IBias}^2(\hat S_0)&=\int_0^{60} \left(\frac{1}{M}\sum_{m=1}^M \hat S^{(m)}_0(u)-S_0(u)\right)^2du,\\
\textrm{IVar}(\hat S_0)&=\frac{1}{M}\sum_{m=1}^M\int_0^{60}\left(\hat S^{(m)}_0(u)-\frac{1}{M}\sum_{m'=1}^M \hat S^{(m')}_0(u)\right)^2du.
\end{align*}
The $\hat S^{(m)}_0$, $m=1,\ldots,M$, represent the estimates for each replication. 
Finally, the total variation between $\hat\lambda_0$ and $\lambda_0$ was also computed for our adaptive ridge estimator. For a given estimate $\hat\lambda_0^{(m)}$, the quantity 
$\textrm{TV}^{(m)}(\hat\lambda_0^{(m)})=\sum_{k=1}^K (c_k-c_{k-1})\mid \exp(\hat a_k)-\exp(a_k)\mid$ 
was calculated in Model M$1$ and the average over all estimates $\textrm{TV}(\hat\lambda_0)=\sum_m \textrm{TV}^{(m)}(\hat\lambda_0^{(m)})/M$ was reported. The results are presented in Tables~\ref{tab:simures1}, ~\ref{tab:simures2} for Model M$1$ and Tables~\ref{tab:simures3},~\ref{tab:simures4} for Model M$2$. {\color{red}Results on the performance of cuts detection are displayed in Tables~\ref{tab:cutdetect} and~\ref{tab:cutdetect2}}. Three different sample sizes ($n=200, 400, 1\,000$) were considered in all models and scenarios, for the midpoint, the ICsurv and the adaptive ridge estimators.

{\color{red}From the simulation results, it is seen that the midpoint estimate has a lower variance than our adaptive ridge estimator both for $\hat\beta$ and $\hat S_0$. However, the midpoint estimator is systematically biased and this bias does not get smaller as the sample size increases. On the other hand, our estimator always has a smaller bias for all scenarios and models and both the bias and the variance decrease as the sample size increases. For example, in Scenario S$1$, Model M$1$, for $n=400$, which corresponds to the sample size of the real data analysis of Section~\ref{sec:data} and to similar proportions of left, interval and right censoring, our estimator exhibits a bias for $\beta=(\log(2),\log(0.8))$ that is $15$ and $4$ times smaller than the bias from the midpoint estimator. For the estimation of $S_0$ the bias of our estimator is more than $40$ times  smaller than the midpoint estimator. The ICsurv estimator shows similar performance as our adaptive ridge estimator in Model M$1$. However in Model M$2$, our estimator has a lower bias than ICsurv but a bigger variance, and a slightly bigger MSE.  In Scenarios S$2$ the effect of adding exact observations is seen to decrease the bias and variance of our estimator. For $n=400$ in Model M$1$, Scenario S$2$ the bias for our estimator of $\beta$ is divided by $4$ and $23$ and the bias for our estimator of $S_0$ is divided by $3$. } 

Finally, the likelihood ratio test approach seems to provide adequate coverage probabilities for $\beta$ especially for $n=400$ and $n=1\,000$, in all scenarios and models. {\color{red}Tables~\ref{tab:cutdetect} and~\ref{tab:cutdetect2} show that, in the piecewise constant baseline scenario (Model M$1$), a majority of one cut is found for $n=200$ and $n=400$, most of the time in the set $[35,55]$ and a majority of two cuts are found for $n=1\,000$, with $44\%$ of chances to detect at least one cut in the set $[10,30]$ and $96\%$ of chances to detect at least one cut in the set $[10,30]$. Due to the wide range of the two visits variables $V_1$ and $V_2$, the algorithm is able at best to detect two cuts under this scenario, and miss most of the time one cut in the set $[35,55]$.} 
More simulations were conducted: scenarios including a cure fraction can be found in the Supplementary Material along with a discussion on computational complexity.

\section{Ankylosis complications for replanted teeth}\label{sec:data}

The method is illustrated on a dental dataset. $322$ patients with $400$ avulsed and replanted permanent teeth were followed-up prospectively in the period from $1965$ to $1988$ at the university hospital in Copenhagen, Denmark. The following replantation procedure was used: the avulsed tooth was placed in saline as soon as the patient was received at the emergency ward. If the tooth was obviously contaminated, it was cleansed with gauze soaked in saline or rinsed with a flow of saline from a syringe. The tooth was replanted in its socket by digital pressure. The patients were then examined at intermittent visits to the dentist. In this study, we focused on a complication called ankylosis characterized by the fusion of the tooth to the bone such that the variable of interest $T$ is the time from replantation of the tooth to ankylosis. This complication may occur if the cells on the root surface is damaged in which case, healing of the periodontal ligament  surrounding the tooth will be impaired, leading to local ingrowth of bone.  Ankylosis cannot be arrested and gradually the root of the tooth will be replaced by bone which will eventually lead to tooth loss. The data are described in great details in~\cite{andreasen1995replant1} {\color{red}and were analysed using our adaptive ridge method in~\cite{lauridsen2019risk}}. 

A total of $28\%$ of the data were left censored, $35.75\%$ were interval censored and $36.25\%$ were right censored. Four covariates were included in the study: the stage of root formation ($72.5\%$ of mature teeth, $27.5\%$ of immature teeth), the length of extra-alveolar storage (mean time is $30.9$ minutes), the type of storage media ($85.25\%$ physiologic, $14.75\%$ non physiologic) and the age of the patient (the mean age for mature teeth is $16.81$ years). There is no need for a cure fraction in this analysis since all different models (non-parametric or regression models) estimated the cure fraction to $0\%$. The adaptive ridge method found four cuts for the baseline hazard at time points $100$, $500$, $800$ and $900$ where the initial grid search was composed of $10$ spaced time points from $0$ to $200$ and then of $100$ spaced time points from $200$ to $2\,000$ ($K_{\text {max}}=40$). The initial grid search was motivated by the data: for $71\%$ of the left and interval-censored data, the right endpoint is lower than $200$.  

\begin{figure}[htb!]
	\begin{minipage}{0.5\textwidth}
		\includegraphics[width=1\textwidth]{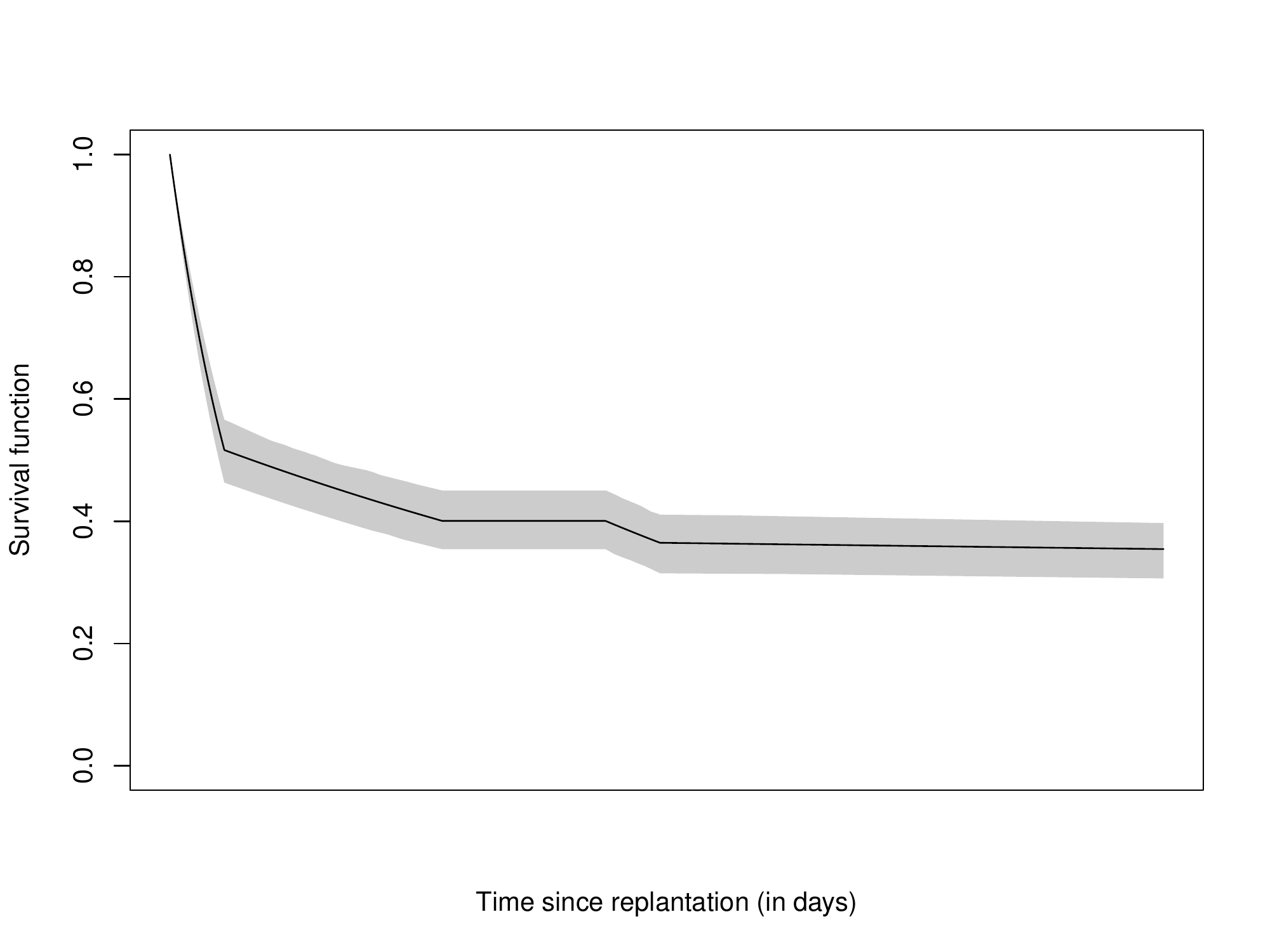}\\
	\end{minipage}
	\begin{minipage}{0.5\textwidth}
		\includegraphics[width=1\textwidth]{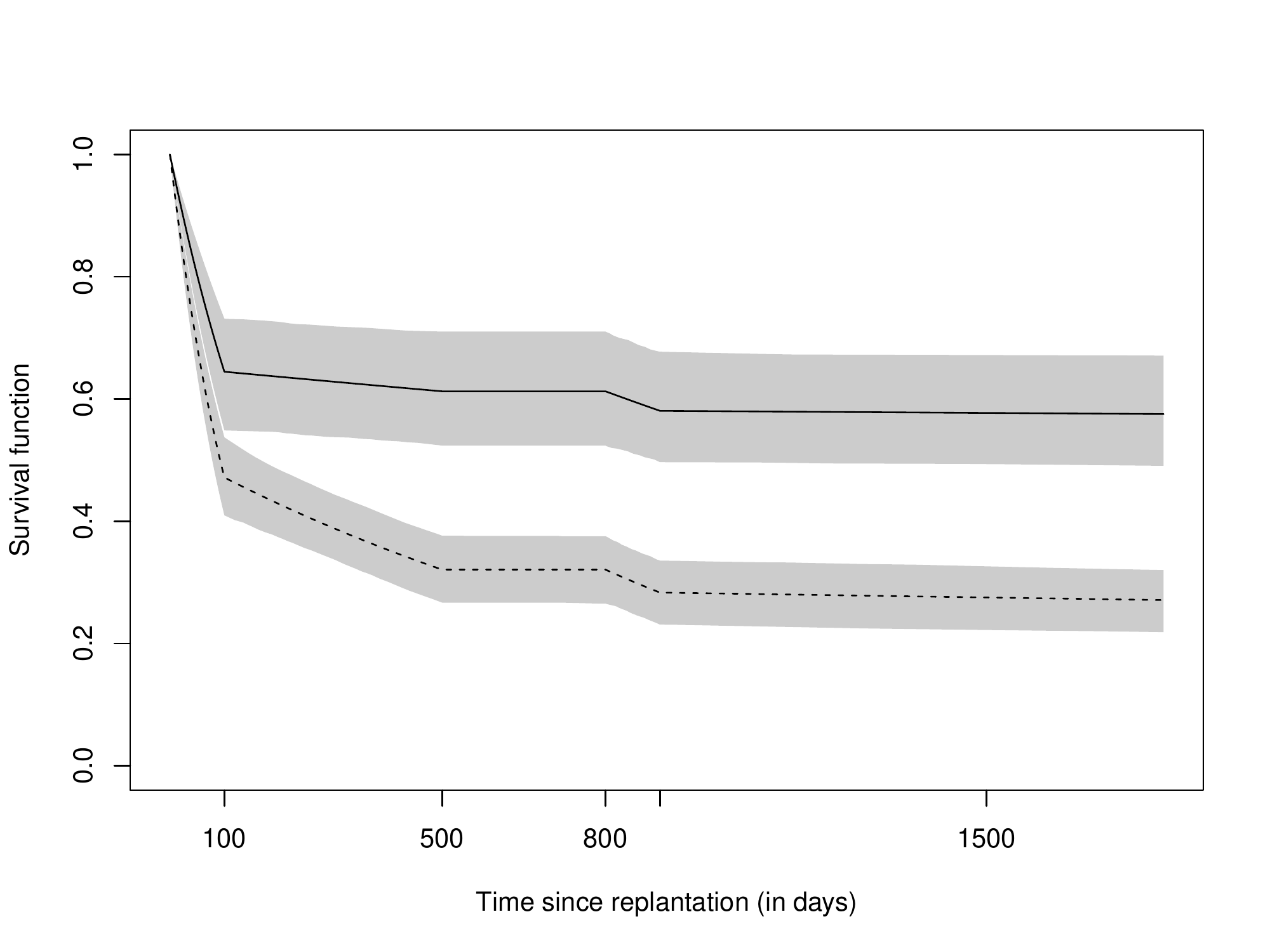}\\
	\end{minipage}
	\caption{On the left panel, estimate of the survival function of time to ankylosis for the whole population. On the right panel, estimates of the survival function for the immature teeth (solid line) and for the mature teeth (dotted lines). Confidence intervals are plotted along the curves in shaded areas using the bootstrap approach.}\label{fig:surv_curve}
\end{figure}

Non-parametric survival estimates were first computed, one for the whole population and two for each subgroup defined by the stage of root formation (see Figure~\ref{fig:surv_curve}). Confidence intervals were also computed using the boostrap method with $500$ replications. These plots illustrate an interesting feature of the adaptive ridge procedure: by selecting a parsimonious set of cuts, the method highlights the different regions of time where the risk of failure varies. There is in particular a very high risk of ankylosis before $100$ days as shown by the very steep survival curve on this time interval. On the global survival curve, the risk of developing ankylosis (one minus the survival function) before $100$ days is estimated to $48.35\%$ $[43.39\%;53.67\%]$. Then the slope of the survival curve decreases from $100$ days to $500$ days, with a risk to develop ankylosis  before $500$ days estimated to $59.94\%$ $[54.96\%;64.57\%]$. The risk of ankylosis after $900$ days is almost null (as shown by the plateau of the survival curve) suggesting that if a patient has not yet developed ankylosis after $900$ days he/she is almost no longer at risk for this complication.  

When looking at the two subgroups defined by stage of root formation we can see that the risk of ankylosis is much higher in the mature group than in the immature group. This is a very interesting result as it confirms the finding from~\cite{andreasen1995replant4} where periodontal ligament healing was seen to be less frequent with advanced stages of root development. From our analysis, it is seen that the risk is in particular higher in the interval $[100,500]$ for the mature group than for the immature group, with ankylosis coming mostly from the mature group in this time range. For the immature group, the risk of developing ankylosis before $100$ days is estimated to $35.54\%$ $[26.85\%;45.13\%]$ and to $52.84\%$ $[46.26\%;59.03\%]$ for the mature teeth. Then the slope of the survival curve decreases from $100$ days to $500$ days, with a risk to develop ankylosis  before $500$ days estimated to $38.74\%$ $[28.97\%;47.62\%]$ for the immature teeth and to $67.92\%$ $[62.36\%;73.31\%]$ for the mature teeth. The risk gets very low after $500$ days for all groups. 

Finally a Cox model was implemented with all the covariates included. Since age shows little variation for immature teeth, this last variable was only included in interaction with the stage of root formation such that the baseline value corresponds to immature teeth and the covariate is defined as age greater than 20 years for mature teeth only. 
The results for the effects of the covariates are shown in Table~\ref{tab:Coxreg}. 
Statistical tests and confidence intervals for each variable were implemented using the log-ratio statistic test as explained in the Supplementary Material (see also Section~\ref{sec:test_CI}). It can be seen that the stage of root formation is highly significant with a two-fold increased risk for mature teeth to develop ankylosis. The storage time is also highly significant with a $1.23$ increase of risk per hour. The type of storage media seems to have no effect on ankylosis and age is not significant even at the $10\%$ level. The baseline hazard values along with their $95\%$ confidence intervals are also displayed in Table~\ref{tab:Coxregbaz}. This hazard corresponds to the risk of immature teeth with non-physiologic type of storage and a storage time of $20$ minutes. We can see how the risk is much higher before $100$ days than at any other time period. Prediction curves for any specific individual can be plotted using these values.


\section{Conclusion}
The estimation method proposed in this paper is very general and allows to deal with a wide range of situations. We first introduced the method for the mixed case of left-censored, interval-censored and right-censored data and we then directly extended it to consider the inclusion of exact observations and a cure fraction. We showed that treating the true event times as unobserved and using the EM algorithm to perform estimation resulted in a diagonal block matrix of the baseline hazard in the piecewise constant Cox model. This is a very interesting feature of our approach since the standard estimation method for this model (see for instance~\cite{sun07}) results in a full rank Hessian matrix, which can pose some serious computational problems for a moderate number of baseline cuts. Moreover, this allowed us to use the {\color{red}$L_0$} penalisation technique developed in~\cite{NuelFrommlet} and~\cite{rippe2012visualization} which was also implemented for exact and right censored data in~\cite{bouaziz17}. Starting from a large grid of baseline cuts this penalisation technique forces two similar adjacent values to be equal. This results in a very flexible model since the location and number of cuts of the baseline are directly determined from the data. As compared to the ICsurv method from~\cite{wang2016flexible}, the EM algorithm is readily applicable without need of a data augmentation step. Even though our cumulative baseline hazard does not result in a smooth function as compared to their spline approach, our method was shown to perform greatly on simulated data and even to outperform the method from~\cite{wang2016flexible} especially in terms of bias of the estimated parameters. It should be mentioned that their method could probably be improved by using an automatic procedure to choose the location and number of knots from the data. However, this is a complicated problem and there is currently no available method that could be directly applied on this estimator (see~\cite{wand2000comparison} for a review on selection methods of knots for spline estimators). 
On the dental dataset we also showed the interesting feature of the adaptive ridge procedure: by detecting the different time regions where the hazard for ankylosis changes, it revealed a very high risk of failure from replantation of the tooth until $100$ days after replantation and a risk near to zero after $900$ days. Finally, theoretical results were also provided for the adaptive ridge estimator. They show that the asymptotic distribution of the parameters can be determined by considering the estimated set of cuts as fixed and by using standard asymptotic likelihood theory for the piecewise constant hazard model.

By use of a logit link we developed the general cure model introduced by~\cite{sy00} and~\cite{peng2000nonparametric}, for interval-censored data. From this model the effect of covariates on the odds of being cured and on the hazard risk of the susceptibles can be assessed. Interestingly, the combination of the piecewise constant baseline hazard and the adaptive ridge procedure produce a very flexible model in this context and avoids the use of arbitrary constraints such as in~\cite{sy00} where the authors had to require that the conditional survival function is set to zero beyond the last event time. 

Another type of heterogeneity could be modelled with the use of frailty models (see~\cite{therneau} for instance). The EM approach for frailty models could then be used as a direct extension of our estimation method. However, it would require to compute the conditional value of the frailty variable given the observed data, a work that is left to future research. Similarly the standard mixture problem where one assumes the population to be composed of two (or more) subgroups with different hazards could be considered (see for instance~\cite{bussy2017c} for this model in a high dimensional setting). The use of the piecewise constant baseline hazard would be crucial for this problem as the model is only identifiable for parametric baselines. The implementation of the adaptive ridge procedure would then result in a very flexible model for this problem.

%

%



\bibliographystyle{abbrv}
\bibliography{biblio}

\begin{table}[!htb]
	\caption{\small Simulation results for the estimation of $\beta$ in Model M$1$ (piecewise constant baseline hazard), for Scenarios S$1$ and S$2$ with $100\%$ of susceptible individuals. S$1$: no exact data, $25\%$ of left-censoring, $52\%$ of interval-censoring, $23\%$ of right-censoring. S$2$: $18\%$ of exact data, $19\%$ of left-censoring, $40\%$ of interval-censoring, $23\%$ of right-censoring.}\label{tab:simures1}
	\npdecimalsign{.}
	\nprounddigits{3}
	\small
	\begin{tabular}{| l| c| cccc| ccc| ccc| }
		\hline 
		&& \multicolumn{4}{c| }{Adaptive Ridge estimate}& \multicolumn{3}{c| }{Midpoint estimate} & \multicolumn{3}{c| }{ICsurv estimate} \\  
		&$n$& \footnotesize Bias($\hat \beta$) & \footnotesize SE($\hat \beta$)& \footnotesize MSE($\hat \beta$) & \footnotesize CP($\hat \beta$)& \footnotesize Bias($\hat \beta$) & \footnotesize SE($\hat \beta$)& \footnotesize MSE($\hat \beta$) & \footnotesize Bias($\hat \beta$) & \footnotesize SE($\hat \beta$) & \footnotesize MSE($\hat \beta$)  \\ 
		\hline 
		S$1$&$200$  &  \numprint{0.03154}& \numprint{0.23525}&  \numprint{0.05633733}& 0.942 & \numprint{-0.17429}  &\numprint{0.18401} &\numprint{0.06423668}&{\color{red}\numprint{0.03754159}} &{\color{red}\numprint{0.2290321}} &{\color{red}\numprint{0.05386509}} \\
		&& \numprint{-0.01049}& \numprint{0.18089}& \numprint{0.03283123}& 0.924  &\numprint{0.05729} &\numprint{0.14092}&\numprint{0.02314059} & {\color{red}\numprint{-0.01675520}} & {\color{red}\numprint{0.1843237}} & {\color{red}\numprint{0.03425597}}\\ 
		\hline 
		&$400$ &  \numprint{0.01160} &\numprint{0.16602}&\numprint{0.0276972} &0.946 & \numprint{-0.17692} &\numprint{0.12710}&\numprint{0.0474551} & {\color{red}\numprint{0.0162697} }&{\color{red}\numprint{0.1604064}}&{\color{red}\numprint{0.02599493}}  \\
		&&\numprint{-0.01367} &\numprint{0.12021}&\numprint{0.01463731}& 0.938 & \numprint{0.05023}&\numprint{0.09609}&\numprint{0.01175634}  &{\color{red}\numprint{-0.01312271}} & {\color{red}\numprint{0.1208187}}& {\color{red}\numprint{0.01476936}}\\ 
		\hline 
		&$1\,000$ &  \numprint{0.00735} & \numprint{0.09861}& \numprint{0.009777955}& 0.948 &\numprint{-0.17135}  &\numprint{0.07467}&\numprint{0.03493643} & {\color{red}\numprint{0.006779654}} &{\color{red}\numprint{0.09565405}}&{\color{red}\numprint{0.009195661}}  \\
		&& \numprint{-0.00275}& \numprint{0.07475}& \numprint{0.005595125}& 0.946  &\numprint{0.05619} &\numprint{0.06203} &\numprint{0.007005037}&{\color{red}\numprint{-0.003161606}} & {\color{red}\numprint{0.07468432}}& {\color{red}\numprint{0.005587743}}\\ 
		\hline \hline 
		S$2$&$200$  &  \numprint{0.03267}& \numprint{0.213265}& \numprint{0.04654929}& 0.945 &\numprint{-0.12813}  &\numprint{0.18143}&\numprint{0.04933414} & & & \\
		&& \numprint{-0.00551}& \numprint{0.16884}& \numprint{0.02853731}& 0.954 &\numprint{0.04548} &\numprint{0.14701}&\numprint{0.02368037} &  & & \\ 
		\hline
		&$400$  &  \numprint{0.00282} &\numprint{0.15280}&\numprint{0.02335579}& 0.947 &\numprint{-0.13758} &\numprint{0.12752}&\numprint{0.03518961} & & &\\
		&& \numprint{-0.00060} &\numprint{0.11904}&\numprint{0.01417088}& 0.952&\numprint{0.045764}&\numprint{0.10443}&\numprint{0.01299997} &  & & \\ 
		\hline &$1\,000$  &  \numprint{0.00560} & \numprint{0.09234}& \numprint{0.008558036}& 0.948 &\numprint{-0.13558}  &\numprint{0.07849}&\numprint{0.02454262} & & &\\
		&& \numprint{0.00230}& \numprint{0.07111}& \numprint{0.005061922}&0.949 &\numprint{0.05072} &\numprint{0.06214}&\numprint{0.006433898} &  & & \\ 
		\hline 
	\end{tabular}
	\npnoround
\end{table}

\begin{table}[htb!]
	\caption{\small Simulation results for the estimation of $S_0$ in Scenarios S$1$ and S$2$ in Model M$1$ (piecewise constant baseline hazard), with $100\%$ of susceptible individuals. S$1$: no exact data, $25\%$ of left-censoring, $52\%$ of interval-censoring, $23\%$ of right-censoring. S$2$: $18\%$ of exact data, $19\%$ of left-censoring, $40\%$ of interval-censoring, $23\%$ of right-censoring.}\label{tab:simures2}
	\npdecimalsign{.}
	\nprounddigits{3}
	\small
	\begin{tabular}{| l| c| ccc| cc| cc| }
		\hline 
		&& \multicolumn{3}{c| }{Adaptive Ridge estimate}& \multicolumn{2}{c| }{Midpoint estimate} & \multicolumn{2}{c| }{ICsurv estimate} \\  
		&$n$& \footnotesize $\mathrm{IBias}^2(\hat S_0)$& \footnotesize $\mathrm{IVar}(\hat S_0)$& \footnotesize $\mathrm{TV}(\hat\lambda_0)$& \footnotesize $\mathrm{IBias}^2(\hat S_0)$&\footnotesize  $\mathrm{IVar}(\hat S_0)$& \footnotesize $\mathrm{IBias}^2(\hat S_0)$& \footnotesize $\mathrm{IVar}(\hat S_0)$  \\ 
		\hline 
		S$1$&$200$  &  \numprint{0.00247} &\numprint{0.26630}&\numprint{0.78424}&  \numprint{0.12400} & \numprint{0.12184}  &{\color{red}\numprint{0.003393097}} &{\color{red}\numprint{0.4378384}} \\
		\hline 
		&$400$ &  \numprint{0.00306} &\numprint{0.13845}&\numprint{0.59976}&\numprint{0.12443} &\numprint{0.06106} &{\color{red}\numprint{0.001790012}}&{\color{red}\numprint{0.2131929}} \\
		\hline 
		&$1\,000$ &  \numprint{0.00241} &\numprint{0.05936}&\numprint{0.41587}& \numprint{0.12566} &\numprint{0.02257} &{\color{red}\numprint{0.001446547}}&{\color{red}\numprint{0.07734946}}  \\
		\hline \hline 
		S$2$&$200$  &  \numprint{0.00148} &\numprint{0.19616}&\numprint{0.64596}& \numprint{0.07414} & \numprint{0.11373} &&\\
		\hline
		&$400$  &  \numprint{0.00137} &\numprint{0.10306}&\numprint{0.48381}&\numprint{0.07434} &\numprint{0.06021} &&\\
		\hline &$1\,000$  &  \numprint{0.00023} &\numprint{0.03812}&\numprint{0.27685}&\numprint{0.07470} &\numprint{0.02221} && \\
		\hline 
	\end{tabular}
	\npnoround
\end{table}

\begin{table}[htb!]
	\caption{\small Simulation results for the estimation of $\beta$ in Model M$2$ (Weibull baseline hazard), for Scenarios S$1$ and S$2$ with $100\%$ of susceptible individuals. S$1$: no exact data, $25\%$ of left-censoring, $52\%$ of interval-censoring, $23\%$ of right-censoring. S$2$: $18\%$ of exact data, $19\%$ of left-censoring, $40\%$ of interval-censoring, $23\%$ of right-censoring.}\label{tab:simures3}
	\npdecimalsign{.}
	\nprounddigits{3}
	\small
	\begin{tabular}{| l| c| cccc| ccc| ccc| }
		\hline 
		&& \multicolumn{4}{c| }{Adaptive Ridge estimate}& \multicolumn{3}{c| }{Midpoint estimate} & \multicolumn{3}{c| }{ICsurv estimate} \\  
		&$n$& \footnotesize Bias($\hat \beta$) & \footnotesize SE($\hat \beta$)& \footnotesize MSE($\hat \beta$) & \footnotesize CP($\hat \beta$) &\footnotesize Bias($\hat \beta$) & \footnotesize SE($\hat \beta$)& \footnotesize MSE($\hat \beta$) & \footnotesize Bias($\hat \beta$) & \footnotesize SE($\hat \beta$) & \footnotesize MSE($\hat \beta$)  \\ 
		\hline 
		S$1$&$200$  &  \numprint{0.02709345}& \numprint{0.5721946}&  \numprint{0.3281408}&\numprint{0.9158317} &\numprint{-0.5961559}  &\numprint{0.1683628} &\numprint{0.38346515}&{\color{red}\numprint{-0.26747915}}&{\color{red}\numprint{0.3068028}} &{\color{red}\numprint{0.1656731}} \\
		&& \numprint{-0.03235672}& \numprint{0.5155291}& \numprint{0.2668173}&0.922 &\numprint{0.1840995} &\numprint{0.1463010}&\numprint{0.05508308} & {\color{red}\numprint{0.09112134}} & {\color{red}\numprint{0.2575091}} & {\color{red}\numprint{0.07461404}}\\ 
		\hline 
		&$400$ &  \numprint{0.02229782} &\numprint{0.4124544}&\numprint{0.1706158} & 0.930 & \numprint{-0.6085846} &\numprint{0.1164921}&\numprint{0.38367558} & {\color{red}\numprint{-0.26291159}}  &{\color{red}\numprint{0.2338650}}&{\color{red}\numprint{0.12381539}}  \\
		&&\numprint{-0.02145597} &\numprint{0.2977516}&\numprint{0.0891164}& 0.934 & \numprint{0.1931046}&\numprint{0.1043176}&\numprint{0.04795498}  &{\color{red}\numprint{0.08684538}} & {\color{red}\numprint{0.1739352}}& {\color{red}\numprint{0.03779557}}\\ 
		\hline 
		&$1\,000$ &  \numprint{0.021127661} & \numprint{0.2059506}& \numprint{0.04286202}& 0.948 &\numprint{-0.6108978}  &\numprint{0.07471135}&\numprint{0.37872229} & {\color{red}\numprint{-0.25110235}} &{\color{red}\numprint{0.1582908}}&{\color{red}\numprint{0.08810837}}  \\
		&& \numprint{0.008977627}& \numprint{0.1699589}& \numprint{0.02896662}& 0.954&\numprint{0.1976972} &\numprint{0.06175499} &\numprint{0.04285981}&{\color{red}\numprint{0.0775192}} & {\color{red}\numprint{0.1115543}}& {\color{red}\numprint{0.01845359}}\\ 
		\hline \hline 
		S$2$&$200$  &  \numprint{-0.08460542}& \numprint{0.2954464}& \numprint{0.09444666}& 0.936&\numprint{-0.5809263}  &\numprint{0.1565465}&\numprint{0.36173771} & & & \\
		&& \numprint{0.01233343}& \numprint{0.2393346}& \numprint{0.05743318}& 0.941 &\numprint{0.1915908} &\numprint{0.1486933}&\numprint{0.05859622} &  & & \\ 
		\hline
		&$400$  &  \numprint{-0.066274178} &\numprint{0.2171575}&\numprint{0.05154964}& 0.942 &\numprint{-0.5821446} &\numprint{0.1150828}&\numprint{0.35207028} & & &\\
		&& \numprint{0.01547789} &\numprint{0.1587355}&\numprint{0.02521516}& 0.950 &\numprint{0.1806218} &\numprint{0.0963573}&\numprint{0.04186261} &  & & \\ 
		\hline &$1\,000$  &  \numprint{-0.04802743} & \numprint{0.1344482}& \numprint{0.02038295}& 0.949 &\numprint{-0.5866512}  &\numprint{0.07231888}&\numprint{0.34933746} & & &\\
		&& \numprint{-0.004265218}& \numprint{0.1032114}& \numprint{0.01067079}& 0.950 &\numprint{0.1898496} &\numprint{0.06132625}&\numprint{0.03976628} &  & & \\ 
		\hline 
	\end{tabular}
	\npnoround
\end{table}

\begin{table}[htb!]
	\caption{\small Simulation results for the estimation of $S_0$ in Scenarios S$1$ and S$2$ in Model M$2$ (Weibull baseline hazard), with $100\%$ of susceptible individuals. S$1$: no exact data, $25\%$ of left-censoring, $52\%$ of interval-censoring, $23\%$ of right-censoring. S$2$: $18\%$ of exact data, $19\%$ of left-censoring, $40\%$ of interval-censoring, $23\%$ of right-censoring.}\label{tab:simures4}
	\npdecimalsign{.}
	\nprounddigits{3}
	\small
	\begin{tabular}{| l| c| cc| cc| cc| }
		\hline 
		&& \multicolumn{2}{c| }{Adaptive Ridge estimate}& \multicolumn{2}{c| }{Midpoint estimate} & \multicolumn{2}{c| }{ICsurv estimate} \\  
		&$n$& \footnotesize $\mathrm{IBias}^2(\hat S_0)$& \footnotesize $\mathrm{IVar}(\hat S_0)$& \footnotesize $\mathrm{IBias}^2(\hat S_0)$&\footnotesize  $\mathrm{IVar}(\hat S_0)$& \footnotesize $\mathrm{IBias}^2(\hat S_0)$& \footnotesize $\mathrm{IVar}(\hat S_0)$  \\ 
		\hline 
		S$1$&$200$  &  \numprint{0.02607795} &\numprint{0.6466932}&  \numprint{1.857449} & \numprint{0.07693194}  &{\color{red}\numprint{0.08227031}} &{\color{red}\numprint{0.2288678}} \\
		\hline 
		&$400$ &  \numprint{0.005032946} &\numprint{0.3908036}&\numprint{1.855556} &\numprint{0.04344094} &{\color{red}\numprint{0.0688195}}&{\color{red}\numprint{0.148202}} \\
		\hline 
		&$1\,000$ &  \numprint{0.00466932} &\numprint{0.1688487}& \numprint{1.931381} &\numprint{0.01455977} &{\color{red}\numprint{0.0500083}}&{\color{red}\numprint{0.05952029}}  \\
		\hline \hline 
		S$2$&$200$  &  \numprint{0.01585109} &\numprint{0.1956695}& \numprint{1.032661} & \numprint{0.08653485} &&\\
		\hline
		&$400$  &  \numprint{0.01041115} &\numprint{0.1044817}&\numprint{1.046383} &\numprint{0.04042707} &&\\
		\hline &$1\,000$  &  \numprint{0.002921194} &\numprint{0.04363108}&\numprint{1.055834} &\numprint{0.0169896} && \\
		\hline 
	\end{tabular}
	\npnoround
\end{table}

\begin{table}[htb!]
\caption{\small {\color{red}Proportions of the number of cuts found by the adaptive ridge algorithm in Scenario S$1$ Model M$1$ (piecewise constant baseline hazard). The true number of cuts is $3$.}}\label{tab:cutdetect}
\begin{tabular}{|c|ccc|}
\hline
Number&\multicolumn{3}{c|}{Proportions found for:}\\
 of cuts&$n=200$&$n=400$&$n=1\,000$\\
\hline\hline
1&{\color{red}0.690}&{\color{red}0.598}&{\color{red}0.400}\\
2& {\color{red}0.288}& {\color{red}0.358}&{\color{red}0.560}\\
  3 & {\color{red}0.020}&{\color{red}0.036}&{\color{red}0.038}\\
  4 & {\color{red}0.002} &{\color{red}0.006}&{\color{red}0.002}\\
  \hline 
\end{tabular}
\end{table}

%
\begin{table}[!htb]
\caption{\small {\color{red}Probabilities that a cut value has been selected by the adaptive ridge algorithm in the sets $[10,30]$ and $[35,55]$ in Scenario S$1$ Model M$1$ (piecewise constant baseline hazard). The true cuts are located at positions $20$, $40$ and $50$}.}\label{tab:cutdetect2}
\begin{tabular}{|c|c|c|c|c|}
	\hline 
	&  & $n=200$ & $n=400$ & $n=1\,000$ \\ 
	\hline 
	Number & 0 & {\color{red}0.718} & {\color{red}0.710} & {\color{red}0.560} \\ 
	of cuts & 1 & {\color{red}0.280} & {\color{red}0.286} & {\color{red}0.434} \\ 
	in $[10,30]$ & 2 & {\color{red}0.020} & {\color{red}0.004} & {\color{red}0.006} \\ 
	\hline \hline 
	Number & 0 & {\color{red}0.198} & {\color{red}0.094} & {\color{red}0.040} \\ 
	of cuts & 1 & {\color{red}0.782} & {\color{red}0.844} & {\color{red}0.860} \\ 
	in $[35,55]$ & 2 & {\color{red}0.020} & {\color{red}0.062} & {\color{red}0.100} \\ 
	\hline 
\end{tabular} 
\end{table}

\begin{table}[htb!]
	\centering
	\begin{tabular}{| c| c| c| c| }
		\hline
		Covariates&HR&$95\%$ CI&p-value\\\hline
		Mature &$2.00$ & $[1.74 ; 2.29]$ & $1.89\times10^{-5}$\\
		Storage time (hours) & $1.23$ & $[1.11 ; 1.34]$ & $0.0017$\\
		Physiologic storage& $0.93$ & $[0.81 ; 1.06]$ & $0.6980$ \\
		Age$>$20 (mature teeth) & $1.27$ & $[0.99 ; 1.61]$ & $0.1272$ \\
		\hline
	\end{tabular}
	\caption{Regression modelling of time to ankylosis on the dental dataset (HR: Hazard Ratio, CI: Confidence Interval). The adaptive ridge found four cuts for the baseline hazard at times $100$, $500$, $800$ and $900$.}\label{tab:Coxreg}
\end{table}

\begin{table}[htb!]
	\centering
	\begin{tabular}{| c| c| c| }
		\hline
		Cuts&$\exp(\hat a_k)\times 10^{3}$&$95\% \text{ CI} \times 10^{3}$\\\hline
		$(0,100]$ &$3.71$ & $[3.19 ; 4.28]$ \\
		$(100,500]$ & $0.39$ & $[0.28 ; 0.52]$ \\
		$(500,800]$& $0.00$ & $[0.00 ; 0.00]$  \\
		$(800,900]$ & $0.62$ & $[0.31 ; 1.07]$  \\
		$(900,+\infty)$ & $0.02$ & $[0.01 ; 0.04]$  \\
		\hline
	\end{tabular}
	\caption{Baseline hazard from the regression modelling of time to ankylosis on the dental dataset (CI: Confidence Interval). This hazard corresponds to the risk of immature teeth with non-physiologic type of storage and a storage time of $20$ minutes.}\label{tab:Coxregbaz}
\end{table}

\clearpage

 \begin{center}
        \vspace*{1cm}

       \LARGE
        \textbf{Supplementary Material}
        \normalsize
        \vspace*{1cm}
    \end{center}

\appendix
\renewcommand{\thesection}{A.\arabic{section}}

\section{Expressions of the statistics $A^{\text{old}}_{k,i}$ and $B^{\text{old}}_{k,i}$}

For $k=1,\ldots,K$, $i=1,\ldots,n$, define
\begin{align*}
A^{\text{old}}_{k,i}&= \frac{\exp\Big(e^{a_{i,k}^{\text{old}}}c_{k-1}+a_{i,k}^{{\text{old}}}-\sum_{j=1}^{k-1}e^{a_{i,j}^{\text{old}}}(c_j-c_{j-1})\Big)J_{k,i}}{S(L_i\mid Z_i,\boldsymbol\theta_{\text{old}})-S(R_i\mid Z_i,\boldsymbol\theta_{\text{old}})}\int_{c_{k-1}\vee L_i}^{c_k\wedge R_i} \exp\big(-e^{a_{i,k}^{\text{old}}}t\big)dt\nonumber\\
&=\exp\Big(-e^{a_{i,k}^{\text{old}}}c_{k-1}\vee L_i\Big)\Big(1-\exp\big(-e^{a_{i,k}^{\text{old}}}(c_k\wedge R_i-c_{k-1}\vee L_i)\big)\Big)\nonumber\\
&\quad \times \frac{\exp\big(e^{a_{i,k}^{\text{old}}}c_{k-1}-\sum_{j=1}^{k-1}e^{a_{i,j}^{\text{old}}}(c_j-c_{j-1})\big)J_{k,i}}{S(L_i\mid Z_i,\boldsymbol\theta_{\text{old}})-S(R_i\mid Z_i,\boldsymbol\theta_{\text{old}})}
\end{align*}
and
\begin{align*}
B^{\text{old}}_{k,i} &= \frac{\exp\Big(e^{a_{i,k}^{\text{old}}}c_{k-1}+a_{i,k}^{\text{old}}-\sum_{j=1}^{k-1}e^{a_{i,j}^{\text{old}}}(c_j-c_{j-1})\Big)J_{k,i}}{S(L_i\mid Z_i,\boldsymbol\theta_{\text{old}})-S(R_i\mid Z_i,\boldsymbol\theta_{\text{old}})}\int_{c_{k-1}\vee L_i}^{c_k\wedge R_i} (t-c_{k-1})\exp(-e^{a_{i,k}^{\text{old}}}t)dt\\
B^{\text{old}}_{k,i} &=\left\{\big(\exp(-a_{i,k}^{\text{old}})+c_{k-1}\vee L_i-c_{k-1}\big)\exp(-e^{a_{i,k}^{\text{old}}}c_{k-1}\vee L_i)\right.\nonumber\\
& \quad \left.-\big(\exp(-a_{i,k}^{\text{old}})+c_k\wedge R_i-c_{k-1}\big)\exp(-e^{a_{i,k}^{\text{old}}}c_k\wedge R_i)\right\}\nonumber\\
&\quad \times \frac{\exp\big(e^{a_{i,k}^{\text{old}}}c_{k-1}-\sum_{j=1}^{k-1}e^{a_{i,j}^{\text{old}}}(c_j-c_{j-1})\big)J_{k,i}}{S(L_i\mid Z_i,\boldsymbol\theta_{\text{old}})-S(R_i\mid Z_i,\boldsymbol\theta_{\text{old}})}\cdot
\end{align*}

The function $Q$ is then expressed as a function of these two statistics (see Section~\ref{sec:fixedcuts} of the main paper).

\section{The Schurr complement}\label{sec:schurr}

The Schurr complement is used to compute the inverse of the Hessian matrix of $Q$, in the case of fixed cuts (Section~\ref{sec:fixedcuts} of the main paper) and of $\ell^{\text{pen}}$, for the adaptive ridge estimator (Section~\ref{sec:ar} of the main paper). It makes use of the special structure of the block matrix corresponding to the second order derivatives with respect to the $a_k$s which is either diagonal (for $Q$) or tri-diagonal (for $\ell^{\text{pen}}$).

Let $\mathcal I(a,\beta)$ be minus the Hessian matrix of $Q$ or $\ell^{\text{pen}}$ for the maximisation problem with respect to $a_1,\ldots,a_L$ and $\beta_1,\ldots,\beta_{d_Z}$. Let $A$ be of dimension $K\times K$, $B$ of dimension $K\times d_Z$ and $C$ be of dimension $d_Z\times d_Z$ such that 
\[\mathcal I(a,\beta)=\begin{pmatrix}
A      & B \\
B^t       & C
\end{pmatrix}\]
Let $U(a,\beta)$ be the score vector of $Q$ or $\ell^{\text{pen}}$ and $b_1$ be the column vector of dimension $K$, $b_2$ be the column vector of dimension $d_Z$ such that $U(a,\beta)=(b_1,b_2)^t$. Using the Schurr complement, we have
\begin{align*}
\mathcal I(a,\beta)^{(-1)}U(a,\beta)=\begin{pmatrix}
A^{-1}b_1- A^{-1}B(C-B^tA^{-1}B)^{-1}(b_2-B^tA^{-1}b_1)  \\
(C-B^tA^{-1}B)^{-1}(b_2-B^tA^{-1}b_1) 
\end{pmatrix}.
\end{align*}
For the inversion of the Hessian matrix of $Q$ and $\ell^{\text{pen}}$, the $K\times K$ matrix $A$ is either diagonal (for $Q$) or a band matrix of bandwidth equal to $1$ (for $\ell^{\text{pen}}$). Its inverse can be efficiently computed using a fast C++ implementation of the LDL algorithm. This is achieved in linear complexity using the R \texttt{bandsolve} package. As a result, the total complexity for the computation of $\mathcal I(a,\beta)^{(-1)}U(a,\beta)$ is of order $\mathcal O(K)$ in the case $K>>d_Z$. 

\section{Score vector and Hessian matrix for the function $Q$ when including exact observations and a cure fraction}

In the presence of exact observations and a cure fraction, the score vector and the Hessian matrix are given from the following formulas:
\begin{align*}
\frac{\partial Q(\boldsymbol\theta\mid \boldsymbol\theta_{\text{old}})}{\partial a_k} & = \sum_{i \text{ not exact}}\pi_i^{\text{old}}\left\{A^{\text{old}}_{k,i}-(c_k-c_{k-1})e^{a_{k}}I(k\neq K)\sum_{l=k+1}^K A^{\text{old}}_{l,i}e^{\beta Z_i}-e^{a_{k}}B^{\text{old}}_{k,i}e^{\beta Z_i}\right\}\\
& \quad +\sum_{i \text{ exact}}\bigg\{O_{i,k}-\exp(a_{k}+\beta Z_i)R_{i,k}\bigg\},\\
\frac{\partial Q(\boldsymbol\theta\mid \boldsymbol\theta_{\text{old}})}{\partial \beta} & = \sum_{i \text{ not exact}}\pi_i^{\text{old}}Z_i\sum_{l=1}^K\left(A^{\text{old}}_{l,i}-\Bigg\{\sum_{j=1}^{l-1}(c_j-c_{j-1})e^{a_{j}}A^{\text{old}}_{l,i}e^{\beta Z_i}+e^{a_{l}}B^{\text{old}}_{l,i}e^{\beta Z_i}\Bigg\}\right)\\
& \quad +\sum_{i \text{ exact}}Z_i\sum_{l=1}^K\bigg\{O_{i,l}-\exp(a_{l}+\beta Z_i)R_{i,l}\bigg\},
\end{align*}
\begin{align*}
\frac{\partial^2 Q(\boldsymbol\theta\mid \boldsymbol\theta_{\text{old}})}{\partial a^2_k} & = -\sum_{i \text{ not exact}}\pi_i^{\text{old}}\left\{(c_k-c_{k-1})e^{a_{k}}I(k\neq K)\sum_{l=k+1}^KA^{\text{old}}_{l,i}e^{\beta Z_i}+e^{a_{k}}B^{\text{old}}_{k,i}e^{\beta Z_i}\right\}\\
&\quad-\sum_{i \text{ exact}}\exp(a_{k}+\beta Z_i)R_{i,k},\\
\frac{\partial^2 Q(\boldsymbol\theta\mid \boldsymbol\theta_{\text{old}})}{\partial \beta^2} & = -\sum_{i \text{ not exact}}\pi_i^{\text{old}}Z_iZ_i^t\sum_{l=1}^K\left(\sum_{j=1}^{l-1}(c_j-c_{j-1})e^{a_{j}}A^{\text{old}}_{l,i}e^{\beta Z_i}+e^{a_{l}}B^{\text{old}}_{l,i}e^{\beta Z_i}\right)\\
&\quad -\sum_{i \text{ exact}}Z_iZ_i^t\sum_{l=1}^K\exp(a_{l}+\beta Z_i)R_{i,l},\\
\frac{\partial^2 Q(\boldsymbol\theta\mid \boldsymbol\theta_{\text{old}})}{\partial a_k\partial \beta} & = -\sum_{i \text{ not exact}}\pi_i^{\text{old}}Z_i\left((c_k-c_{k-1})e^{a_{k}}I(k\neq K)\sum_{l=k+1}^{K}A^{\text{old}}_{l,i}e^{\beta Z_i}+e^{a_{k}}B^{\text{old}}_{k,i}e^{\beta Z_i}\right),\\
&\quad-\sum_{i \text{ exact}}Z_i\exp(a_{k}+\beta Z_i)R_{i,k}.
\end{align*}

\section{Full regularisation path on a simulated dataset}\label{sec:full_reg_path}

\begin{figure}[!htb]
\begin{tabular}{cc}
\includegraphics[width=0.47\textwidth,height=0.5\textwidth]{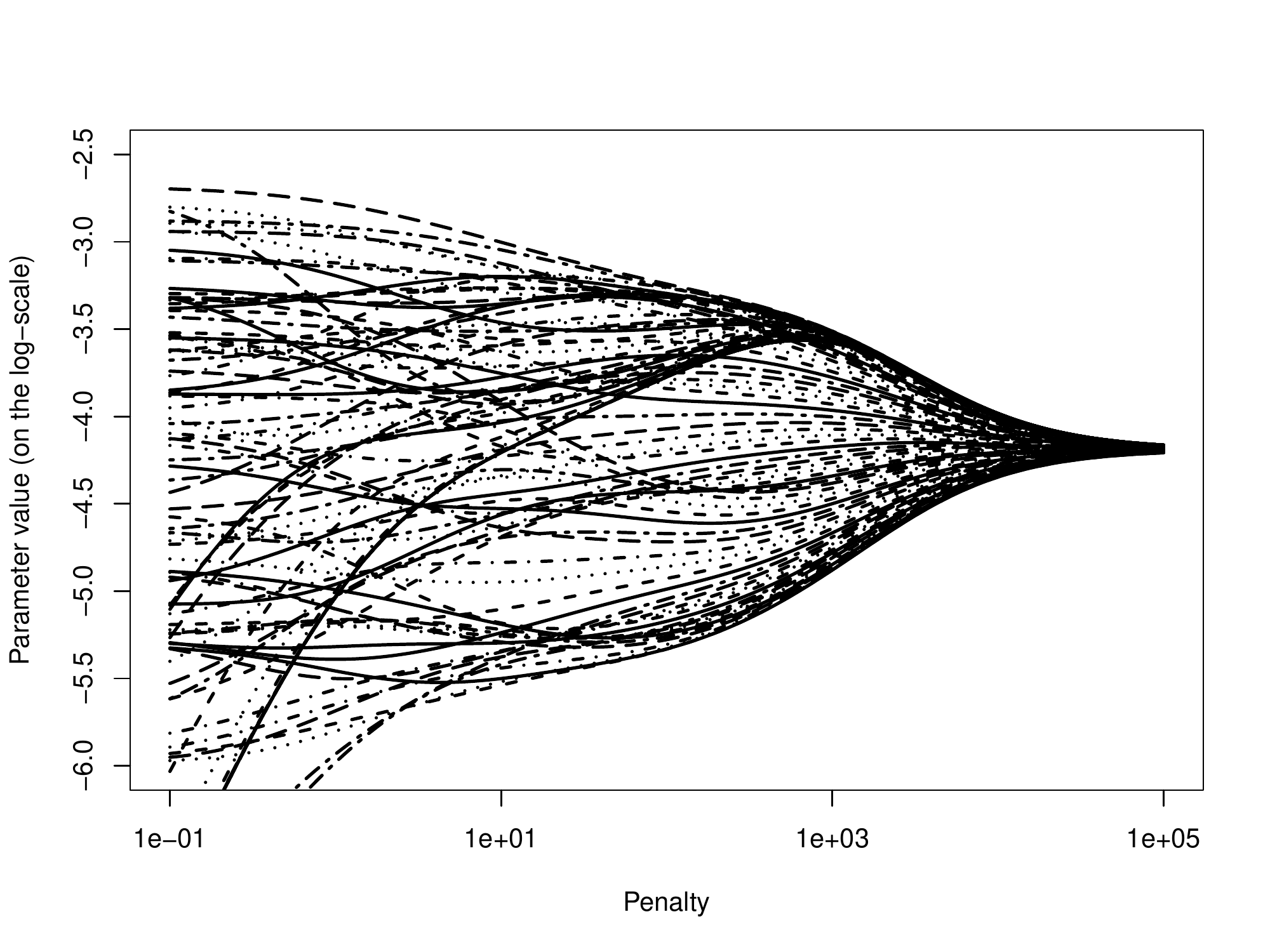}&\includegraphics[width=0.47\textwidth,height=0.5\textwidth]{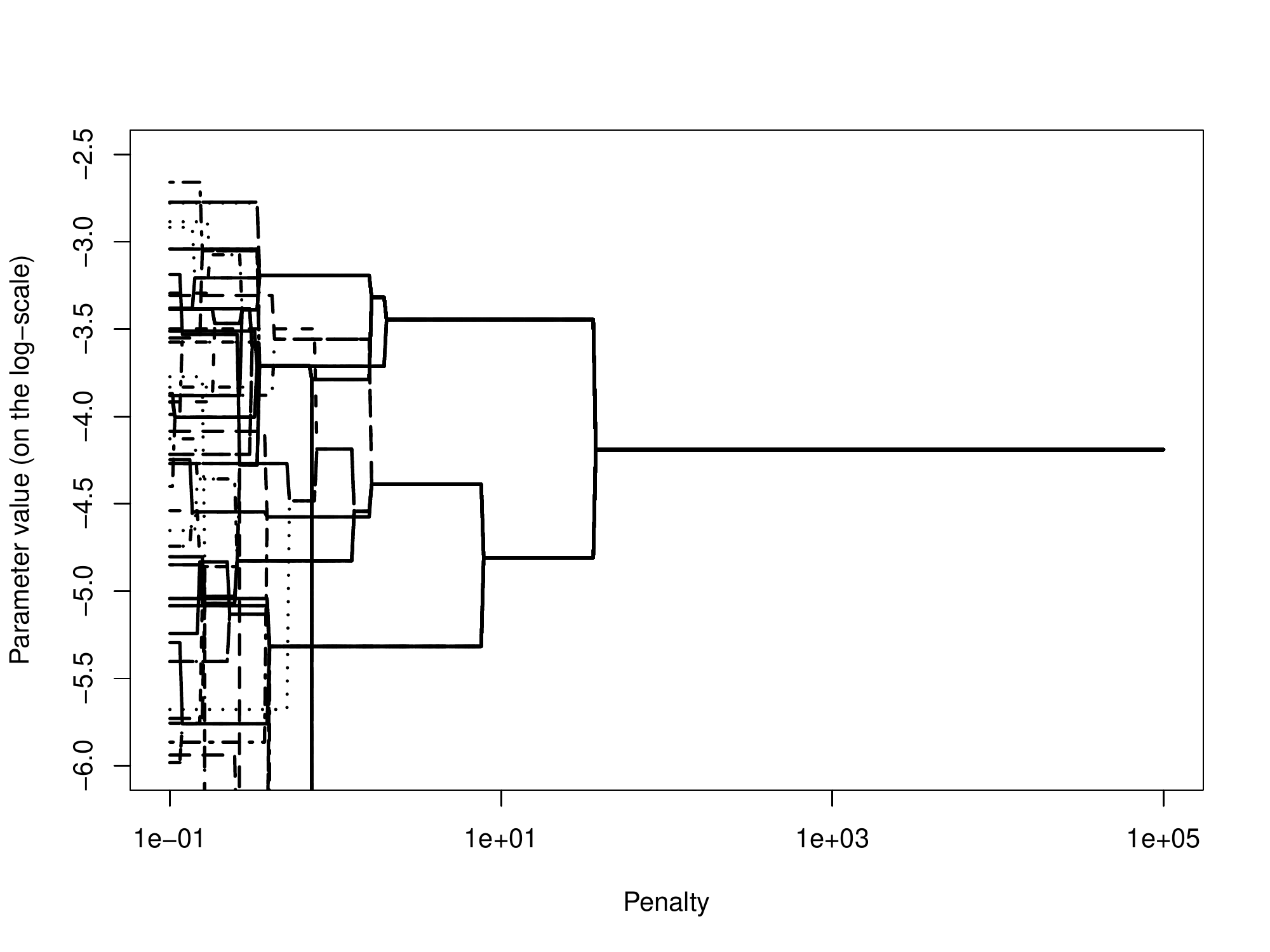}
\end{tabular}
\caption{Regularization path for the ridge on the left panel and for the adaptive ridge on the right panel. The $x$-axis represents the penalty value and the $y$-axis represents the estimated values of the $a_k$'s.}
\label{fig:reg_path}
\end{figure}

\begin{figure}[!htb]
\begin{tabular}{cc}
\includegraphics[width=0.47\textwidth,height=0.5\textwidth]{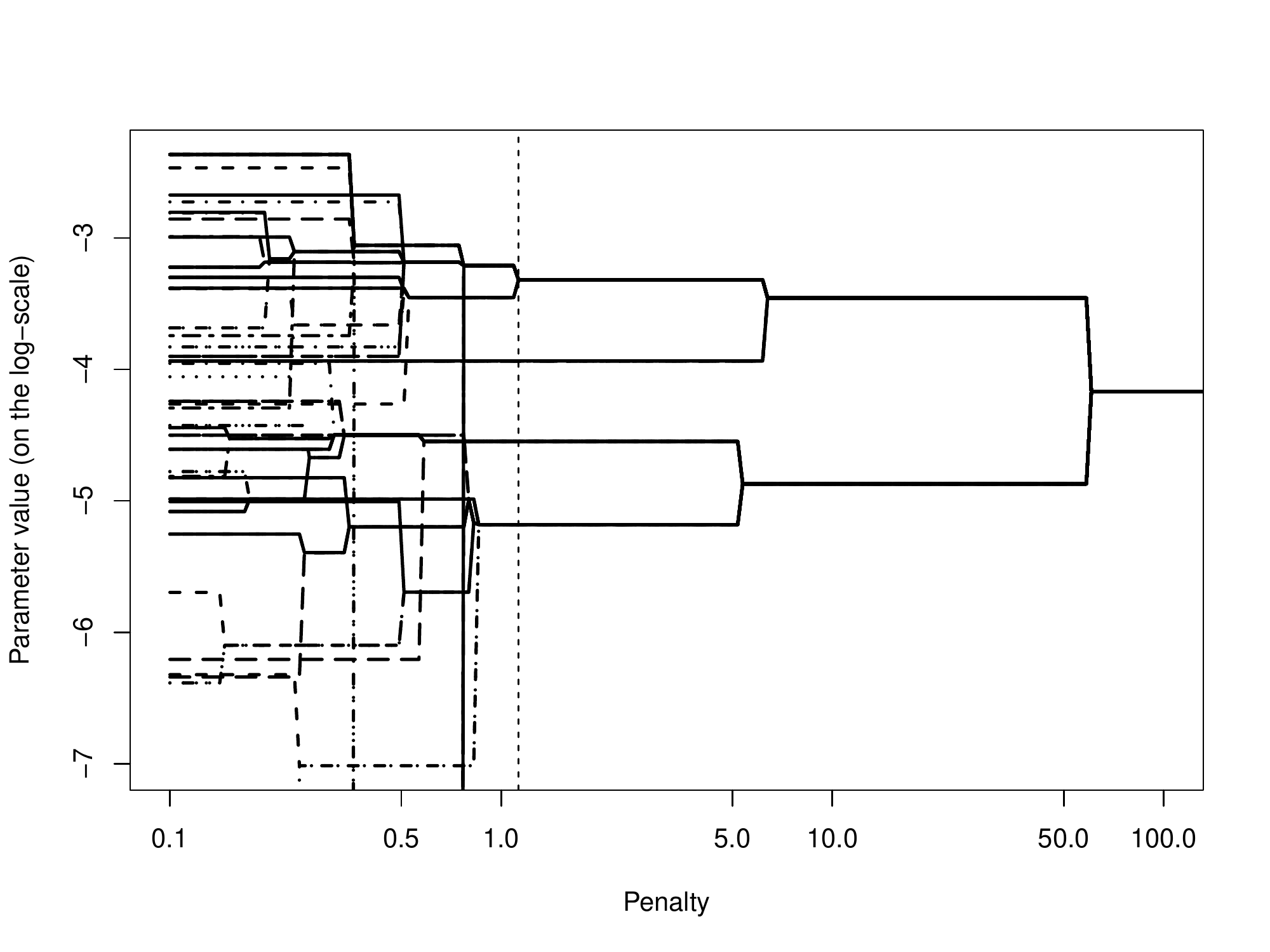}&\includegraphics[width=0.47\textwidth,height=0.5\textwidth]{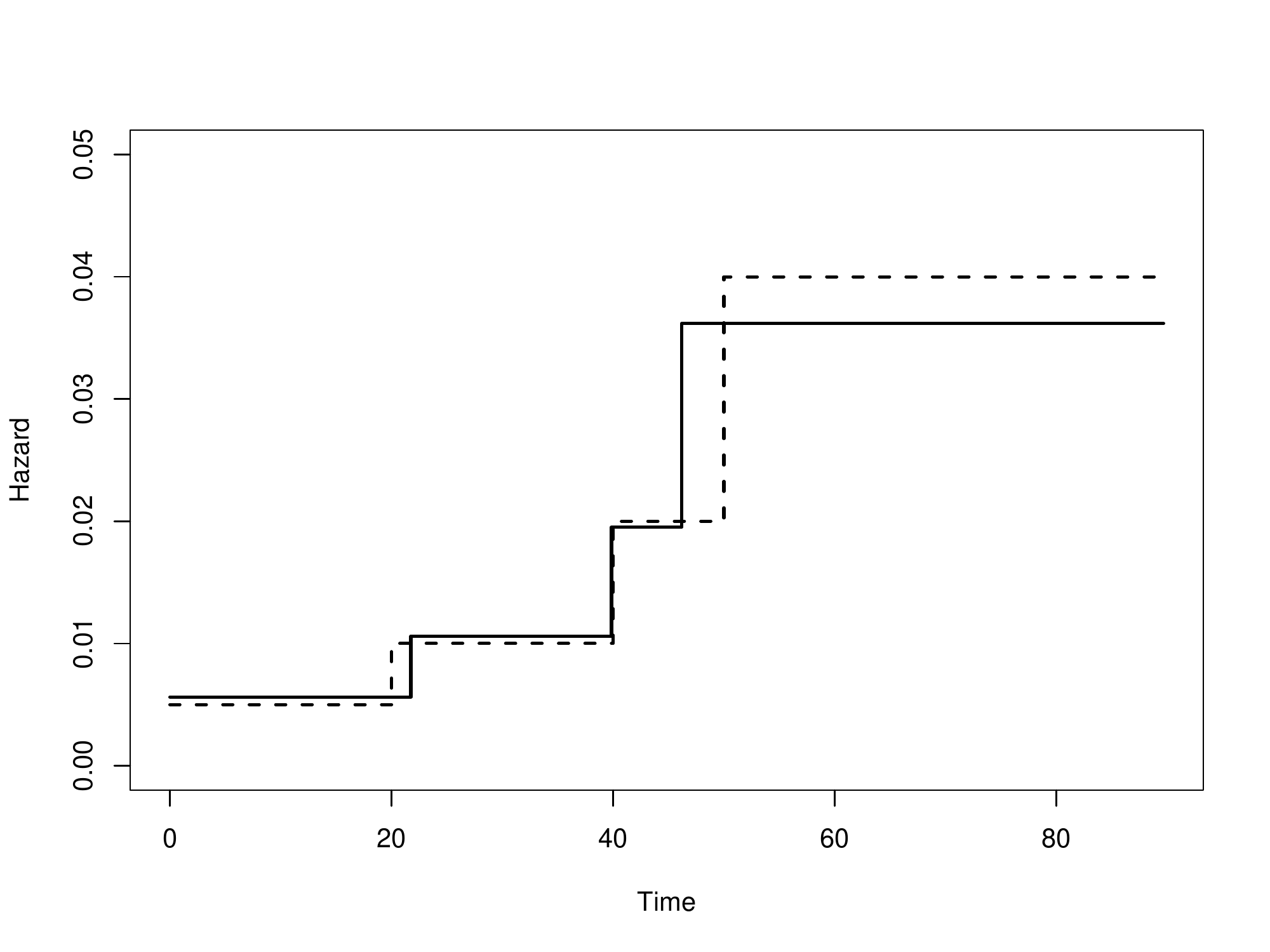}
\end{tabular}
\caption{Regularization path for the adaptive-ridge on the left panel. The estimated set of cuts using the BIC is shown as a vertical dotted line. The resulting piecewise constant hazard estimator is shown on the right panel as a solid line. The dotted line represents the true hazard.}
\label{fig:haz_BIC}
\end{figure}

We illustrate in this section the full regularisation path of the algorithm. As explained in Section~\ref{sec:ar} of the main paper the algorithm consists of the detection of the set of cuts from the penalised estimator combined with the non-penalised estimator using this estimated set of cuts. We consider one sample generated from Model M$1$, Scenario S$1$ of Section~\ref{sec:simu} of the main paper in the absence of covariates and we estimate the hasard function using both the ridge and the adaptive ridge algorithm. More precisely, the first algorithm uses the weights $\hat w_k$ equal to $1$ while the second algorithm iteratively updates the $\hat w_k$ using Equation~\eqref{eq:ARalgo} of the main paper. A set of penalty is chosen, on the log scale, as the set of $200$ equally spaced values ranging from $\log(0.1)$ to $\log(10\,000)$. Figure~\ref{fig:reg_path} displays the regularisation path for the ridge on the left and for the adaptive ridge on the right where the $y$-axis represents the values of the estimated $a_k$'s for each penalty value of the $x$-axis. We clearly see that the ridge procedure produces a smooth estimation and the adaptive ridge procedure provides a selection of the cuts along with an estimated piecewise constant hazard. Both estimators converge toward the same constant model as pen tends to infinity. Figure~\ref{fig:haz_BIC} shows the resulting estimated hazard from the adaptive ridge procedure after selection of the cuts using the BIC. On the left panel it is seen that the BIC chooses a model with three cuts and four values of $a_k$'s. On the right panel we see that, on this sample, the adaptive ridge estimator follows closely the true value of the hazard.

\section{Proof of Theorem 5.1 of the main document}
\textsc{ Proof of 1.}  

For this proof, we only consider the initial fixed set of cuts $\{c_1,\ldots,c_K\}$. In order to avoid confusion, we denote by $\boldsymbol{\theta^{\dag}}=(a_1^{\dag},\ldots,a_K^{\dag},\beta^{*})$ the true parameter using this set of cuts. This means that there might exist several $k$'s for which $a_k^{\dag}=a_{k+1}^{\dag}$. Note that removing the equal consecutive values of $a_k^{\dag}$ will yield $\boldsymbol{\theta^{*}}$. In the following, we will prove that $\boldsymbol{\hat{\theta}}\to\boldsymbol{\theta^{\dag}}$ in probability.

For interval-censored, left or right-censored data, the full likelihood function can be written as 
\begin{align*}
\tilde{\mathrm{L}}_n^{\text{obs}}(\boldsymbol \theta) &= \prod_{i=1}^n (f_{L,R,\delta}(L_i,R_i,1))^{\delta_i}(f_{L,R,\delta}(L_i,R_i,0))^{1-\delta_i},
\end{align*}
where $f_{L,R,\delta}(L_i,R_i,1), f_{L,R,\delta}(L_i,R_i,0)$ represent the joint density of the mixed distribution $(L,R,\delta)$ respectively evaluated at $(L_i,R_i,1)$ and $(L_i,R_i,0)$. It is then seen that $f_{L,R,\delta}(L_i,R_i,1)=\mathbb P[\delta=1\mid L=L_i,R=R_i,Z_i,\boldsymbol\theta] f_{L,R,Z}(L_i,R_i,Z_i)$ where $f_{L,R,Z}$ represents the joint density of $(L,R,Z)$ and $\mathbb P[\delta=1\mid L=L_i,R=R_i,Z_i,\boldsymbol\theta]=(S(L_i\mid Z_i,\boldsymbol{\theta})-S(R_i\mid Z_i,\boldsymbol{\theta}))^{\delta_i}$ under the independent censoring assumption. The same kind of reasoning holds for $f_{L,R,\delta}(L_i,R_i,0)$ such that
\begin{align*}
\tilde{\mathrm{L}}_n^{\text{obs}}(\boldsymbol \theta) &= \prod_{i=1}^n (S(L_i\mid Z_i,\boldsymbol{\theta})-S(R_i\mid Z_i,\boldsymbol{\theta}))^{\delta_i} (S(L_i\mid Z_i,\boldsymbol{\theta}))^{1-\delta_i}f_{L,R,Z}(L_i,R_i,Z_i),\\
&=\prod_{i=1}^n g_{\boldsymbol\theta}(L_i,R_i,Z_i),
\end{align*}
where $g_{\boldsymbol\theta}(L_i,R_i,Z_i):=(S(L_i\mid Z_i,\boldsymbol{\theta})-S(R_i\mid Z_i,\boldsymbol{\theta}))f_{L,R,Z}(L_i,R_i,Z_i)$ with the slight abuse of notation $S(R_i\mid Z_i,\boldsymbol{\theta})=0$ if $R_i=\infty$ (for a right-censored observation). The above equation shows that the full likelihood is simply the observed likelihood $\mathrm{L}_n^{\text{obs}}(\boldsymbol \theta)$ of Section~\ref{sec:EM} of the main document multiplied by the quantity $f_{L,R,Z}(L_i,R_i,Z_i)$ which does not depend on $\boldsymbol{\theta}$. In case of exact observations, the full likelihood can be rewritten as:
\begin{align*}
\tilde{\mathrm{L}}_n^{\text{obs}}(\boldsymbol \theta) &= \prod_{i \text{ not exact}} g_{\boldsymbol\theta}(L_i,R_i,Z_i)\prod_{i \text{ exact}} f(L_i\mid Z_i,\boldsymbol\theta).
\end{align*}
It should be noted that $g_{\boldsymbol\theta}(L_i,R_i,Z_i)$ and $f(L_i\mid Z_i,\boldsymbol\theta)$ are densities. For $g_{\boldsymbol\theta}$, write
\begin{align*}
\iiint_{l\neq r} g_{\boldsymbol\theta}(l,r,z)dldrdz&=\mathbb E_{\boldsymbol\theta}\Big[I(L_i\neq R_i)\mathbb E_{\boldsymbol\theta}[S(L_i\mid Z_i,\boldsymbol{\theta})-S(R_i\mid Z_i,\boldsymbol{\theta})\mid L,R,Z]\Big]\\
&=\iiint \mathbb P[T\in (l,r)\mid L=l,R=r,Z=z,\boldsymbol \theta)f_{L,R,Z}(l,r,z)dldrdz.
\end{align*}
From the independent censoring assumption, $\mathbb P[T\in (l,r)\mid L=l,R=r,Z=z,\boldsymbol \theta)]=1$ and consequently $g_{\boldsymbol\theta}$ is a density.

Now the penalised estimator defined in~\eqref{eq:penobs} of the main document verifies $\boldsymbol{\hat\theta}=\argmax_{\boldsymbol\theta} \ell_n^{\text{pen}}(\boldsymbol\theta)$, where 
\begin{align*}
 \ell_n^{\text{pen}}(\boldsymbol\theta)&=\left\{\ell_n(\boldsymbol\theta)-\frac{\text{pen}}{2n}\sum_{k=1}^{K-1}\hat w_k^{(1)}(a_{k+1}-a_k)^2\right\},
\end{align*}
with $\ell_n(\boldsymbol\theta)=\log(\tilde{\mathrm{L}}_n^{\text{obs}}(\boldsymbol \theta))/n$. We introduce $\ell(\boldsymbol\theta)=\mathbb E_{\boldsymbol{\theta^{\dag}}}[I(L_i\neq R_i)\log(g_{\boldsymbol\theta}(L_i,R_i,Z_i))]+\mathbb E_{\boldsymbol{\theta^{\dag}}}[I(L_i=R_i)\log(f(L_i\mid Z_i,\boldsymbol\theta))]$ and we write:
\begin{align*}
\left |  \ell_n^{\text{pen}}(\boldsymbol\theta)- \ell(\boldsymbol\theta)\right|\leq \left |  \ell_n(\boldsymbol\theta)- \ell(\boldsymbol\theta)\right| + \frac{\text{pen}}{2n}\sum_{k=1}^{K-1}\hat w_k^{(1)}(a_{k+1}-a_k)^2.
\end{align*}
The two terms on the right-hand side of the equation converge toward $0$ in probability: the first one from the law of large numbers, and the second one from the consistency of $\hat w_k^{(1)}$ and the condition $\text{pen}/n \to 0$. 

Then, from Jensen inequality,
\begin{align*}
\mathbb E_{\boldsymbol{\theta^{\dag}}}\left[-I(L_i\neq R_i)\log\left(\frac{g_{\boldsymbol\theta}(L_i,R_i,Z_i)}{g_{\boldsymbol{\theta^{\dag}}}(L_i,R_i,Z_i)}\right)\right]&\geq \mathbb -\log\left(\mathbb E_{\boldsymbol{\theta^{\dag}}}\left[I(L_i\neq R_i)\frac{g_{\boldsymbol\theta}(L_i,R_i,Z_i)}{g_{\boldsymbol{\theta^{\dag}}}(L_i,R_i,Z_i)}\right]\right)\\
&\geq -\log\left(\iiint_{l\neq r} \frac{g_{\boldsymbol\theta}(l,r,z)}{g_{\boldsymbol{\theta^{\dag}}}(l,r,z)}g_{\boldsymbol{\theta^{\dag}}}(l,r,z)dldrdz\right)=0.
\end{align*}
The same reasoning applies to $\mathbb E_{\boldsymbol{\theta^{\dag}}}[I(L_i=R_i)\log(f(L_i\mid Z_i,\boldsymbol\theta)/f(L_i\mid Z_i,\boldsymbol{\theta^{\dag}}))]$ which proves that $\ell(\boldsymbol\theta)\leq \ell(\boldsymbol{\theta^{\dag}})$ for all $\boldsymbol\theta$. To conclude, we have proved that $\left |  \ell_n(\boldsymbol\theta)- \ell(\boldsymbol\theta)\right|\to 0$ in probability, with $\boldsymbol{\hat\theta}=\argmax_{\boldsymbol\theta} \ell_n^{\text{pen}}(\boldsymbol\theta)$ and $\boldsymbol{\theta^{\dag}}=\argmax_{\boldsymbol\theta} \ell(\boldsymbol{\theta})$. The concavity of $\ell_n^{\text{pen}}(\boldsymbol\theta)$ yields that $\boldsymbol{\hat\theta}\to\boldsymbol{\theta^{\dag}}$ in probability.\\

\noindent\textsc{ Proof of 2. and 3.}

We start by working on the true set of cuts $\mathcal A^*$. 
We need to define the estimator $\boldsymbol{\hat{\hat\theta}}_{\mathcal A^*}$, that is our estimator using the true set of cuts. In particular we need to define the value of $\hat{\hat a}_{k,\mathcal A^*}$ on each interval $c^*_{k-1}<t\leq c^*_k$. As a matter of fact, for a given $n$ the sets $\mathcal A_n$ and $\mathcal A^*$ might be different and therefore some $\hat{\hat a}_{k,\mathcal A^*}$ might not exist. We set:
\begin{align*}
\exp(\hat{\hat a}_{k,\mathcal A^*})=\hat{\hat \lambda}_{0,\mathcal A_n}(c^*_{k-1}).
\end{align*}
This definition is arbitrary and any value of $t\in (c^*_{k-1},c^*_k]$ could be taken for $\hat{\hat \lambda}_{0,\mathcal A_n}(t)$. We now also define $\ell_{n,\mathcal A^*}(\boldsymbol \theta)=\log(\mathrm{L}_{n,\mathcal A^*}^{\text{obs}}(\boldsymbol \theta))$ the observed log-likelihood defined using the true set of cuts $\mathcal A^*$. From a Taylor expansion, we have:
\begin{align*}
\nabla_{\boldsymbol\theta} \ell_{n,\mathcal A^*}(\boldsymbol{\hat{\hat\theta}}_{\mathcal A^*})&=\nabla_{\boldsymbol\theta} \ell_{n,\mathcal A^*}(\boldsymbol{\theta^*})+(\boldsymbol{\hat{\hat\theta}}_{\mathcal A^*}-\boldsymbol{\theta^*})^t \nabla^2_{\boldsymbol\theta} \ell_{n,\mathcal A^*}(\boldsymbol{\tilde\theta}_{\mathcal A^*}),
\end{align*}
where $\boldsymbol{\tilde\theta}_{\mathcal A^*}$ is on the line segment between $\boldsymbol{\hat{\hat\theta}}_{\mathcal A^*}$ and $\boldsymbol{\theta^*}$. As a consequence,
\begin{align}\label{eq:taylor}
\sqrt n(\boldsymbol{\hat{\hat\theta}}_{\mathcal A^*}-\boldsymbol{\theta^*})^t &=-(\nabla^2_{\boldsymbol\theta} \ell_{n,\mathcal A^*}(\boldsymbol{\tilde\theta}_{\mathcal A^*})/n)^{-1}(\nabla_{\boldsymbol\theta} \ell_{n,\mathcal A^*}(\boldsymbol{\theta^*})-\nabla_{\boldsymbol\theta} \ell_{n,\mathcal A^*}(\boldsymbol{\hat{\hat\theta}}_{\mathcal A^*}))\frac{1}{\sqrt n}\cdot
\end{align}
From the result in 1. of this theorem, $\boldsymbol{\hat{\hat\theta}}_{\mathcal A^*}\to \boldsymbol{\theta^*}$ in probability, and thus $\nabla^2_{\boldsymbol\theta} \ell_{n,\mathcal A^*}(\boldsymbol{\tilde\theta}_{\mathcal A^*})/n-\nabla^2_{\boldsymbol\theta} \ell_{n,\mathcal A^*}(\boldsymbol{\theta^*})/n$ converges to $0$ in probability and $-\nabla^2_{\boldsymbol\theta} \ell_{n,\mathcal A^*}(\boldsymbol{\tilde\theta}_{\mathcal A^*})/n\to -\mathbb E[\nabla_{\boldsymbol \theta}^2 h^*_{\boldsymbol\theta}(L_i,R_i,Z_i))|_{\boldsymbol\theta=\boldsymbol{\theta}^*}]=\Sigma$ in probability.

The key to the proof is now to show that $\nabla_{\boldsymbol\theta} \ell_{n,\mathcal A^*}(\boldsymbol{\hat{\hat\theta}}_{\mathcal A^*})/\sqrt n$ converges to $0$ in probability. We denote by $\boldsymbol{\hat{\theta}}_{\mathcal A^*}$ the estimator that maximises $\ell_{n,\mathcal A^*}(\boldsymbol \theta)$. Noticing that $\nabla_{\boldsymbol\theta} \ell_{n,\mathcal A^*}(\boldsymbol{\hat{\theta}}_{\mathcal A^*})=0$ we have
\begin{align}\label{eq:0term}
\nabla_{\boldsymbol\theta} \ell_{n,\mathcal A^*}(\boldsymbol{\hat{\hat\theta}}_{\mathcal A^*})/\sqrt n&=\sqrt n(\boldsymbol{\hat{\hat\theta}}_{\mathcal A^*}-\boldsymbol{\hat{\theta}}_{\mathcal A^*})^t \nabla^2_{\boldsymbol\theta} \ell_{n,\mathcal A^*}(\boldsymbol{\tilde\theta}_{\mathcal A^*})/n,
\end{align}
where $\boldsymbol{\tilde\theta}_{\mathcal A^*}$ is on the line segment between $\boldsymbol{\hat{\hat\theta}}_{\mathcal A^*}$ and $\boldsymbol{\hat{\theta}}_{\mathcal A^*}$. Since $\boldsymbol{\hat{\theta}}_{\mathcal A^*}\to \boldsymbol{\theta^*}$ and $\boldsymbol{\hat{\hat\theta}}_{\mathcal A^*}-\boldsymbol{\hat{\theta}}_{\mathcal A^*}\to 0$ in probability, we can prove as previously that $\nabla^2_{\boldsymbol\theta}\ell_{n,\mathcal A^*}(\boldsymbol{\tilde\theta}_{\mathcal A^*})/n\to \Sigma$ in probability.

We now work on the initial set of cuts  $\{c_1,\ldots,c_K\}$ and we define $\boldsymbol{\hat\theta^{\dag}}$, the estimator $\boldsymbol{\hat{\theta}}_{\mathcal A^*}$ that is defined on $\{c_1,\ldots,c_K\}$ (this is always possible since $\mathcal A^* \subset \{c_1,\ldots,c_K\}$). We need to prove that $\sqrt n(\boldsymbol{\hat\theta}-\boldsymbol{\hat\theta^{\dag}})^t$ converges to $0$ in probability which will imply that $\sqrt n(\boldsymbol{\hat{\hat\theta}}_{\mathcal A^*}-\boldsymbol{\hat{\theta}}_{\mathcal A^*})^t$ converges to $0$ in probability. Introduce the function:
\begin{align*}
\psi_n(u,v):=\ell_n(\boldsymbol{\hat\theta^{\dag}}+(u,v)/\sqrt n)- \ell_n(\boldsymbol{\hat\theta^{\dag}})-\frac{\text{pen}}{2n}\sum_{k=1}^{K-1}\hat w_k^{(1)}(V(\hat a^{\dag}_k+u_k/\sqrt n)-V(\hat a^{\dag}_k)),
\end{align*}
where $(u,v)=(u_1,\ldots,u_K,v_1,\ldots,v_{d_Z})$ is a row vector of dimension $(K+d_Z)$ and $V(a_k)=(a_{k+1}-a_k)^2$. For 
\begin{align*}
(\hat u,\hat v)=\argmin_{u,v} \psi_n(u,v),
\end{align*}
we have $\boldsymbol{\hat a}=\boldsymbol{\hat a^{\dag}}+\hat u/\sqrt n$ and $\boldsymbol{\hat\beta}=\boldsymbol{\hat\beta^{\dag}}+\hat v/\sqrt n$, that is $\hat u=\sqrt n (\boldsymbol{\hat a}-\boldsymbol{\hat a^{\dag}})$ and $\hat v=\sqrt n(\boldsymbol{\hat\beta}-\boldsymbol{\hat\beta^{\dag}})$. We now study the limit of $\psi_n$. First of all,
\begin{align*}
\ell_n(\boldsymbol{\hat\theta^{\dag}}+(u,v)/\sqrt n)- \ell_n(\boldsymbol{\hat\theta^{\dag}})&=\frac{(u,v)}{\sqrt n}\nabla_{\theta} \ell_n(\boldsymbol{\hat\theta^{\dag}})+\frac {1}{2n} (u,v) \nabla^2_{\theta} \ell_n(\boldsymbol{\hat\theta^{\dag}})(u,v)^t+o_{\mathbb P}(1),
\end{align*}
where the $o_{\mathbb P}(1)$ is obtained from the law of large numbers applied to the partial derivatives of order three of $\ell_n(\boldsymbol{\tilde\theta}_n)$, for a $\boldsymbol{\tilde\theta}_n$ on the line segment between $\boldsymbol{\hat\theta^{\dag}}$ and $(u,v)/\sqrt n$.
By definition, $\boldsymbol{\hat\theta^{\dag}}$ maximises $\ell_n$ and therefore $\nabla_{\theta} \ell_n(\boldsymbol{\hat\theta^{\dag}})=0.$ By the law of large numbers, $\frac {1}{2n} (u,v) \nabla^2_{\theta} \ell_n(\boldsymbol{\hat\theta^{\dag}})(u,v)^t$ converges in probability toward $\frac {1}{2} (u,v) \nabla^2_{\theta} \ell(\boldsymbol{\theta^{\dag}})(u,v)^t=-\frac {1}{2} (u,v)\Sigma(u,v)^t$. 
Secondly,
\begin{align*}
V(\hat a^{\dag}_k+u_k/\sqrt n)-V(\hat a^{\dag}_k)&=\frac{2}{\sqrt n} (\hat a^{\dag}_{k+1}-\hat a^{\dag}_{k})(u_{k+1}-u_k)+\frac{(u_{k+1}-u_k)^2}{n}.
\end{align*}
Since $\hat w_k^{(1)}\to ((a^{\dag}_{k+1}-a^{\dag}_{k})^2+\varepsilon^2)^{-1}$, $\hat a^{\dag}_{k+1}-\hat a^{\dag}_{k}\to a^{\dag}_{k+1}-a^{\dag}_{k}$ in probability and
\begin{align*}
\left|\frac{a^{\dag}_{k+1}-a^{\dag}_{k}}{(a^{\dag}_{k+1}-a^{\dag}_{k})^2+\varepsilon^2}\right |<1,
\end{align*}
we see that $V(\hat a^{\dag}_k+u_k/\sqrt n)-V(\hat a^{\dag}_k)\to 0$ in probability. To summarise we have shown that $\psi_n(u,v)\to -\frac {1}{2} (u,v)\Sigma(u,v)^t$ in probability. Since $\Sigma$ is a positive definite matrix, $-\frac {1}{2} (u,v)\Sigma(u,v)^t$ is minimal for $(u,v)=(0,0)$. This proves that $\sqrt n(\boldsymbol{\hat\theta}-\boldsymbol{\hat\theta^{\dag}})^t$ converges to $0$ in probability.

Going back to Equations~\eqref{eq:taylor} and~\eqref{eq:0term}, and from the asymptotic normality of $\nabla_{\boldsymbol\theta} \ell_{n,\mathcal A^*}(\boldsymbol{\theta^*})/\sqrt n$ using the Central Limit Theorem, we finally obtain:
\begin{align*}
\sqrt n(\boldsymbol{\hat{\hat\theta}}_{\mathcal A^*}-\boldsymbol{\theta^*})^t &=-(\nabla^2_{\boldsymbol\theta} \ell_{n,\mathcal A^*}(\boldsymbol{\tilde\theta}_{\mathcal A^*})/n)^{-1}(\nabla_{\boldsymbol\theta} \ell_{n,\mathcal A^*}(\boldsymbol{\theta^*}))\frac{1}{\sqrt n}+o_{\mathbb P}(1)
\longrightarrow\Sigma^{-1} \mathcal N(0,\Sigma),
\end{align*}
in distribution. This concludes the proof.

\section{Extended simulation study for the piecewise constant hazard model: two scenarios that include exact observations and a cure fraction}

We consider two new scenarios which include a proportion of non-susceptible individuals. For the susceptibles, the data include left, interval and right-censored observations along with a proportion of exact observations. The model is defined by Equations~\eqref{eq:CoxCure} and~\eqref{eq:probcure} of the main paper with a logistic link for the probability of being cured. In both scenarios, the $Z$ covariate, $\beta$ coefficient and $\lambda_0$ baseline function are all generated as in the simulation section of the main paper. The $X$ covariate is of dimension $d_X=2$ (including the intercept) and follows a Bernoulli distribution with parameter $0.8$. In Scenario S$3$, $\gamma=(\log(2.35),\log(2))^t$ and in Scenario S$4$, $\gamma=(\log(0.8),\log(2))^t$. These values yield an average number of susceptible individuals $\mathbb E[p(X)]$ respectively equal to $80\%$ and $58\%$. Among the susceptibles, both scenarios correspond to a proportion of $18\%$ of exact observations, $19\%$ of left observations, $40\%$ of interval-censored observations and $23\%$ of right-censored observations. The results are presented in Table~\ref{tab:simures3sup}. Only our adaptive ridge estimator has been implemented for these two scenarios. The $\gamma$ estimator is initialised to $0$ in the EM algorithm.

A slight deterioration of the variance estimation of $\hat \beta$ and $\hat \lambda_0$ is seen when a cure fraction is included and the degree of deterioration increases as the proportion of cured gets bigger. On the other hand the bias of the parameter estimates is similar with or without the cure fraction. In the presence of a cure fraction, the $\gamma$ parameter is less accurately estimated as compared to the $\beta$ parameter both in terms of bias and variance. Nevertheless the results show that as the sample size increases the bias and variance of $\hat\gamma$ get smaller with a bias very close to $0$ for a sample size equal to $1\,000$. The estimation performance of $\mathbb E[p(X)]$ was also investigated by computing the average value of $\sum_i \hat p(X_i)/n$ for all generated samples where $\hat p(X)$ is defined as in Equation~\eqref{eq:probcure} of the main paper with $\gamma$ replaced by $\hat\gamma$. For example, in Scenario S$4$ we found a bias and empirical standard error (SE) equal for $n=200$ to $0.057$ (SE $=0.064$), for $n=400$ to $0.046$ (SE $=0.044$) and for $n=1\,000$ to $0.033$ (SE $=0.028$).

More simulations were conducted. In particular, the cure model without covariates for the cure fraction was also implemented in Scenario S$1$, Model M$1$ of the main paper such that the parameters to be estimated are $\boldsymbol{\theta}=(a_1, \ldots, a_L,\beta,p)$ with the true value of $p$ equal to $1$. In replications of samples of size $400$, it was seen that the model estimated the proportion of susceptibles $p$ to a value greater than $0.99$ in $98\%$ of cases and the lowest value on the $500$ replications for the estimation of $p$ was equal to $0.95$. This highlights the very high specificity of our model in terms of detecting a cure fraction. It shows that our model does not tend to overestimate the proportion of cured when the population is homogeneous, which is a very important feature of the estimation method. On the other hand, a scenario identical to Scenario S$1$, Model M$1$ but with a true proportion of susceptibles equal to $p=0.7$ was also considered. In replications of samples of size $400$, the estimator of $p$ was equal to $0.712$ on average and only $0.5\%$ of the estimates where greater than $0.99$. This suggests in turn a high sensitivity of our model to detect heterogeneity in interval censored data. 

\section{Computational cost of the adaptive ridge algorithm}


The complexity for the inversion of the Hessian of $\ell$ is of order $\mathcal{O}(K)$, in the case $K>>d_X+d_Z$ (see Section~\ref{sec:schurr} in the Supporting Information about the Schurr complement). However, for a given penalty, it should be noted that the global algorithm for maximising $Q$ or $\ell^{\text{pen}}$ consists of an EM algorithm with a Newton-Raphson procedure at each step. As a consequence, in the simulations and for the dental dataset a Generalised Expectation Maximisation (GEM) algorithm (see~\cite{dempster1977maximumbis}) is used instead of the standard EM where, as soon as the value of $Q$ or $\ell^{\text{pen}}$ increases, the Newton-Raphson procedure is stopped. This results in computing only a few steps of the Newton-Raphson algorithm (very often only one step is needed). As the EM algorithm is usually very slow to reach convergence the \texttt{turboEM} R package with the \texttt{squareEM} option is used to accelerate the procedure (see for instance~\cite{varadhan2008simplebis}). Finally, the algorithm must be iterated for the whole sequence of penalties. In order to evaluate the global computational cost, numerical experiments were conducted which showed that, for a maximum of $K_{\textrm{max}}$ initial cuts, the total complexity of the whole procedure is of order $\mathcal{O}(n K^{1/2}_{\textrm{max}})$. 

More specifically, the computation time for the method was evaluated on replicated samples for the three sample sizes $n=200, 400, 1\,000$ and for different values of the maximal number of initial cuts: $K_{\textrm{max}}=18, 40, 80$. We estimated the implementation of the whole method with $200$ penalty values to $0.0016 \times nK_{\textrm{max}}^{1/2}$ minutes. For example, for $n=400,K_{\textrm{max}}=40$ the whole program takes $4$ minutes, for $n=400,K_{\textrm{max}}=80$ it takes $5.7$ minutes, for $n=1\,000,K_{\textrm{max}}=40$ it takes $10.12$ minutes and for $n=1\,000,K_{\textrm{max}}=80$ it takes $14.3$ minutes. These values are given as an indication of the algorithmic complexity and should be considered with caution as the implementation has not been optimised. In particular, computation of the $A_{k,i}^{\text{old}}$ and $B_{k,i}^{\text{old}}$ terms could be improved by computing the set of values $(c_k\wedge R_i, c_{k-1}\vee L_i)$ such that $(L_i,R_i)\cap(c_{k-1},c_k)\neq \emptyset$ more efficiently in C++. Also the non-penalised MLE is implemented for each selection of cuts. For small penalty values, the set of selected cuts can be quite large and the \texttt{turboEM} R package has trouble to converge in these cases. For very large set of selected cuts it often does not converge at all and the algorithm is stopped after $200$ iterations. This procedure could be greatly improved by only implementing the MLE for reasonable sets of cuts.

Finally, it should be noted that the adaptive ridge procedure needs only to be implemented once on the dataset, in order to detect the set of cuts. Then given this set of cuts, the piecewise-constant hazard model is much faster to compute. For example in Scenario S$1$ from the main paper with three cuts, the computation time of the piecewise-constant hazard maximum likelihood model is on average respectively equal to $1.13$, $1.80$ and $3.33$ seconds for $n=200,400, 1\,000$.

\section{The likelihood ratio approach to construct confidence intervals}

As shown in Section~\ref{sec:test_CI}, statistical inference in our model reduces to a fully parametric problem since, after selection of the cuts, one can consider these cuts as fixed and the asymptotic distribution of the final estimator is identical to the asymptotic distribution one would get if the true cuts were initially provided. 

Statistical tests are implemented from the likelihood ratio test which is based on the observed likelihood $\mathrm{L}_n^{\mathrm{obs}}$. Let $\boldsymbol{\theta}=(\theta_1,\theta_2)$ with $\theta_1$ of dimension $d$. To test the null hypothesis $\mathrm{H}_0:\theta_1=\theta_0$, with $\theta_0$ known, one can use the test statistic  $-2\log(\mathrm{L}_n^{\mathrm{obs}}(\theta_0,\hat \theta_2)/\mathrm{L}_n^{\mathrm{obs}}(\hat \theta_1,\hat \theta_2))$ which follows a chi-squared distribution with $d$ degrees of freedom from standard likelihood theory. 
Confidence intervals can also be constructed from the likelihood ratio statistic. 
Let us assume that $\boldsymbol{\theta}=(\theta_1,\theta_2)$ with $\theta_1$ of dimension $1$ and consider the test $\mathrm{H}_0:\theta_1=\theta_0$ versus $\mathrm{H}_1:\theta_1\neq \theta_0$. The $1-\alpha$ confidence interval level of the parameter $\theta_1$ will be determined by the set of values $\theta_0$ such that the previous test is not significant at the significance level $\alpha$. 
Note that the p-value of the test is defined by (with a slight abuse of notation for the realisation of the test statistic)
\begin{align*}
\mathbb P\left[\chi^2(1)>-2\log\left(\frac{\mathrm{L}_n^{\mathrm{obs}}(\theta_0,\hat{\theta}_2)}{\mathrm{L}_n^{\mathrm{obs}}(\hat{\theta}_1,\hat{\theta}_2)}\right)\right],
\end{align*}
and the test is non-significant if this value is greater than $\alpha$. Let $q_{\chi^2}^{1-\alpha}$ be the $1-\alpha$ quantile of the $\chi^2(1)$ distribution. The bounds of the confidence intervals can therefore be determined by resolving the equation
\begin{align}\label{eq:CI_LR}
\log(\mathrm{L}_n^{\mathrm{obs}}(\theta_0,\hat{\theta}_2))+\frac 12 q_{\chi^2}^{1-\alpha}-\log(\mathrm{L}_n^{\mathrm{obs}}(\hat{\theta}_1,\hat{\theta}_2))=0,
\end{align}
with respect to $\theta_0$. This equation has two solutions and since it is clear that $\theta_0=\hat\theta_1$ is part of the confidence interval (the p-value equals one for this value), a grid search can be performed using for example the {\bf \texttt {uniroot}} package with the two starting intervals $[\hat \theta_1-c;\hat\theta_1]$ and $[\hat \theta_1;\hat\theta_1+c]$, where $c$ is a positive constant. This constant can be chosen arbitrarily large and should satisfy that the left-hand side of Equation~\eqref{eq:CI_LR} is of opposite sign for $\theta_0=\hat \theta_1-c$ and $\theta_0=\hat\theta_1+c$. See~\cite{zhou2015empiricalbis} for more details about the likelihood ratio test approach for constructing confidence intervals.

A more classical method for deriving confidence intervals 
can be based on the normal approximation of the model parameter obtained from Theorem~\ref{theo:result}. It requires to compute the Hessian matrix of the observed log-likelihood. The details for this approach are given in the next section. 

\section{Score vector and Hessian matrix for the observed log-likelihood}

Computation of the Hessian matrix of the observed log-likelihood $\partial^2\log(\mathrm{L}_n^{\mathrm{obs}}(\boldsymbol{\theta}))/\partial\boldsymbol{\theta}^2$ evaluated at $\boldsymbol{\theta}=\hat\theta$ 
can be done by direct calculation or by using the following relationship which makes use of the complete likelihood $\mathrm{L}_n$ (see~\cite{louis1982findingbis}):
\begin{align}\label{eq:score_comp}
\frac{\partial \log\left(\mathrm{L}_n^{\mathrm{obs}}(\boldsymbol{\theta})\right)}{\partial \boldsymbol{\theta}}&=\mathbb{E}\left[\frac{\partial \log\left(\mathrm{L}_n(\boldsymbol{\theta})\right) }{\partial \boldsymbol{\theta}}\,\middle| \,\text{data},\boldsymbol{\theta}\right].
\end{align}
In the above equation, the Hessian can be computed based on the complete likelihood by 
taking the derivative of the right-hand side of the equation with respect to $\boldsymbol{\theta}$. 
For simplicity, we assume that all individuals are susceptibles. Then,
\begin{align*}
\log\left(\mathrm{L}_n(\boldsymbol{\theta})\right)&=\sum_{i \text{ not exact}} \sum_{k=1}^K I(c_{k-1}<T_i\leq c_k)\Big(a_{i,k}-\sum_{j=1}^k e^{a_{i,j}}(T_i\wedge c_j-c_{j-1})\Big),\\
&\quad +\sum_{i \text{ exact}}\sum_{k=1}^K \big\{O_{i,k}a_{i,k}-\exp(a_{i,k})R_{i,k}\big\}\\
\frac{\partial \log\left(\mathrm{L}_n(\boldsymbol{\theta})\right) }{\partial a_k}&=\sum_{i\text{ not exact}}^n \Big\{ I(c_{k-1}<T_i\leq c_k)-\sum_{l=k}^K I(c_{l-1}<T_i\leq c_l)e^{a_{i,k}}(T_i\wedge c_k-c_{k-1})\Big\},\\
&\quad + \sum_{i \text{ exact}} \big\{O_{i,k}-\exp(a_{i,k})R_{i,k}\big\}\\
\frac{\partial \log\left(\mathrm{L}_n(\boldsymbol{\theta})\right) }{\partial \beta}&=\sum_{i=1}^n \sum_{l=1}^K I(c_{l-1}<T_i\leq c_l)Z_i\Big(1-\sum_{j=1}^l e^{a_{i,j}}(T_i\wedge c_j-c_{j-1})\Big)\\
&\quad + \sum_{i \text{ exact}}\sum_{l=1}^KZ_i\big\{O_{i,l}-\exp(a_{i,l})R_{i,l}\big\}.
\end{align*}
We now need to take the expectation conditionally on the data of the last two equations. This will involve the quantities
\begin{align*}
\mathbb P[c_{k-1}<T_i\leq c_k\mid\text{data},\boldsymbol{\theta}]&=\frac{S(c_{k-1}\vee L_i\mid Z_i,\boldsymbol{\theta})-S(c_{k}\wedge R_i\mid Z_i,\boldsymbol{\theta})}{S( L_i\mid Z_i,\boldsymbol{\theta})-S(R_i\mid Z_i,\boldsymbol{\theta})},
\end{align*}
and 
\begin{align*}
& \mathbb E[I(c_{k-1}<T_i\leq c_k)T_i\mid\text{data},\boldsymbol{\theta})]\\
&= J_{k,i}\int_{c_{k-1}\vee L_i}^{c_k\wedge R_i}t \exp\Big(a_{i,k}-\sum_{j=1}^{k}e^{a_{i,j}}(t\wedge c_j-c_{j-1})\Big)dt\times\frac{1}{S(L_i\mid Z_i,\boldsymbol\theta)-S(R_i\mid Z_i,\boldsymbol\theta)},
\end{align*}
\begin{align*}
&=\left\{\big(\exp(-a_{i,k})+c_{k-1}\vee L_i\big)\exp(-e^{a_{i,k}}c_{k-1}\vee L_i)-\big(\exp(-a_{i,k})+c_k\wedge R_i\big)\exp(-e^{a_{i,k}}c_k\wedge R_i)\right\}\nonumber\\
&\quad \times \frac{\exp\big(e^{a_{i,k}}c_{k-1}-\sum_{j=1}^{k-1}e^{a_{i,j}}(c_j-c_{j-1})\big)J_{k,i}}{S(L_i\mid Z_i,\boldsymbol\theta)-S(R_i\mid Z_i,\boldsymbol\theta)}\cdot
\end{align*}
Calculation of the right-hand side of Equation~\eqref{eq:score_comp} is now straightforward.  
We first separate exact and non exact observations in the following way:
\begin{align*}
\frac{\partial\log(\mathrm{L}_n^{\mathrm{obs}}(\boldsymbol{\theta}))}{\partial\boldsymbol{\theta}}&=\sum_{i \text{ not exact}}\frac{\partial\mathrm{L}_{i,1}^{\mathrm{obs}}(\boldsymbol{\theta})}{\partial\boldsymbol{\theta}} + \sum_{i \text{ exact}}\frac{\partial\mathrm{L}_{i,2}^{\mathrm{obs}}(\boldsymbol{\theta})}{\partial\boldsymbol{\theta}}.
\end{align*}
For the non-exact observations, we introduce 
\begin{align*}
C_{i,k}(\boldsymbol{\theta})&=\frac{S(c_{k-1}\vee L_i\mid Z_i,\boldsymbol{\theta})-S(c_{k}\wedge R_i\mid Z_i,\boldsymbol{\theta})}{S( L_i\mid Z_i,\boldsymbol{\theta})-S(R_i\mid Z_i,\boldsymbol{\theta})},\\
D_{i,k}(\boldsymbol{\theta})&=J_{k,i}\left\{\big(\exp(-a_{i,k})+c_{k-1}\vee L_i\big)\exp(-e^{a_{i,k}}c_{k-1}\vee L_i)\right.\\
&\quad\left.-\big(\exp(-a_{i,k})+c_k\wedge R_i\big)\exp(-e^{a_{i,k}}c_k\wedge R_i)\right\} \frac{\exp\big(e^{a_{i,k}}c_{k-1}-\sum_{j=1}^{k-1}e^{a_{i,j}}(c_j-c_{j-1})\big)}{S(L_i\mid Z_i,\boldsymbol\theta)-S(R_i\mid Z_i,\boldsymbol\theta)},
\end{align*}
such that
\begin{align*}
\frac{\partial\mathrm{L}_{i,1}^{\mathrm{obs}}(\boldsymbol{\theta})}{\partial a_k}&=C_{i,k}(\boldsymbol{\theta})-e^{a_{i,k}}\Big(D_{i,k}(\boldsymbol{\theta})-c_{k-1}C_{i,k}(\boldsymbol{\theta})\Big)-e^{a_{i,k}}(c_k-c_{k-1})\sum_{l=k+1}^K C_{i,l}(\boldsymbol{\theta}),\\
\frac{\partial\mathrm{L}_{i,1}^{\mathrm{obs}}(\boldsymbol{\theta})}{\partial \beta}&=Z_i\bigg\{C_{i,k}(\boldsymbol{\theta})-C_{i,k}(\boldsymbol{\theta})\sum_{j=1}^{k-1}e^{a_{i,j}}(c_j-c_{j-1})-e^{a_{i,k}}\Big(D_{i,k}(\boldsymbol{\theta})-c_{k-1}C_{i,k}(\boldsymbol{\theta})\Big)\bigg\}.
\end{align*}
For the exact observations we have
\begin{align*}
\frac{\partial\mathrm{L}_{i,2}^{\mathrm{obs}}(\boldsymbol{\theta})}{\partial a_k}&=O_{i,k}-\exp(a_{k}+\beta Z_i)R_{i,k},\\
\frac{\partial\mathrm{L}_{i,2}^{\mathrm{obs}}(\boldsymbol{\theta})}{\partial \beta}&=Z_i\sum_{l=1}^K\bigg\{O_{i,l}-\exp(a_{l}+\beta Z_i)R_{i,l}\bigg\}.
\end{align*}
For the Hessian matrix $\partial^2\log(\mathrm{L}_n^{\mathrm{obs}}(\boldsymbol{\theta}))/\partial\boldsymbol{\theta}^2$, we first compute 

\begin{align*}
\frac{\partial S(c_{k-1}\!\vee L_i\mid Z_i,\boldsymbol{\theta})}{\partial a_k}&=-\left(L_i I_k(L_i)+c_k I(L_i>c_k)\right)e^{a_{i,k}}S(L_i\mid Z_i,\boldsymbol{\theta}),\\
\frac{\partial S(c_{k-1}\!\vee L_i\mid Z_i,\boldsymbol{\theta})}{\partial \beta}&=-Z_i\sum_{l=1}^K (c_l\wedge c_{k-1}\!\vee L_i-c_{l-1})I(c_{l-1}\leq c_{k-1}\vee L_i)e^{a_{i,k}}S(c_{k-1}\!\vee L_i\mid Z_i,\boldsymbol{\theta}),\\
\frac{\partial S(c_{k}\wedge R_i\mid Z_i,\boldsymbol{\theta})}{\partial a_k}&=-(c_k\wedge R_i-c_{k-1})e^{a_{i,k}}S(c_{k}\wedge R_i\mid Z_i,\boldsymbol{\theta})  I(R_i\geq c_{k-1}),\\
\frac{\partial S(c_{k}\wedge R_i\mid Z_i,\boldsymbol{\theta})}{\partial \beta}&=-Z_i\sum_{l=1}^K (c_l\wedge c_{k}\wedge R_i-c_{l-1})I(c_{l-1}\leq c_{k}\wedge R_i)e^{a_{i,k}}S(c_{k}\wedge R_i\mid Z_i,\boldsymbol{\theta}),\\
\frac{\partial S(L_i\mid Z_i,\boldsymbol{\theta})}{\partial a_k}&=-(c_k\wedge L_i-c_{k-1}) e^{a_{i,k}}S(L_i\mid Z_i,\boldsymbol{\theta})I(L_i\geq c_{k-1}),\\
\frac{\partial S(L_i\mid Z_i,\boldsymbol{\theta})}{\partial \beta}&=-Z_i\sum_{l=1}^K(c_l\wedge L_i-c_{l-1}) e^{a_{i,l}}S(L_i\mid Z_i,\boldsymbol{\theta})I(L_i\geq c_{l-1}),\\
\frac{\partial S(R_i\mid Z_i,\boldsymbol{\theta})}{\partial a_k}&=-(c_k\wedge R_i-c_{k-1}) e^{a_{i,k}}S(R_i\mid Z_i,\boldsymbol{\theta})I(R_i\geq c_{k-1}),\\
\frac{\partial S(R_i\mid Z_i,\boldsymbol{\theta})}{\partial \beta}&=-Z_i\sum_{l=1}^K(c_l\wedge R_i-c_{l-1}) e^{a_{i,l}}S(R_i\mid Z_i,\boldsymbol{\theta})I(R_i\geq c_{l-1}),
\end{align*}
such that calculation of the partial derivatives of $C_{i,k}(\boldsymbol{\theta})$ are calculated from the formulas
\begin{align*}
\frac{\partial C_{i,k}(\boldsymbol{\theta})}{\partial a_k}&=\frac{\partial S(c_{k-1}\vee L_i\mid Z_i,\boldsymbol{\theta})/\partial a_k-\partial S(c_{k}\wedge R_i\mid Z_i,\boldsymbol{\theta})/\partial a_k}{S( L_i\mid Z_i,\boldsymbol{\theta})-S(R_i\mid Z_i,\boldsymbol{\theta})}\\
&\quad-C_{i,k}(\boldsymbol{\theta})\frac{\partial S(L_i\mid Z_i,\boldsymbol{\theta})/\partial a_k-\partial S(R_i\mid Z_i,\boldsymbol{\theta})/\partial a_k}{S( L_i\mid Z_i,\boldsymbol{\theta})-S(R_i\mid Z_i,\boldsymbol{\theta})},\\
\frac{\partial C_{i,k}(\boldsymbol{\theta})}{\partial \beta}&=\frac{\partial S(c_{k-1}\vee L_i\mid Z_i,\boldsymbol{\theta})/\partial \beta-\partial S(c_{k}\wedge R_i\mid Z_i,\boldsymbol{\theta})/\partial \beta}{S( L_i\mid Z_i,\boldsymbol{\theta})-S(R_i\mid Z_i,\boldsymbol{\theta})}\\
&\quad-C_{i,k}(\boldsymbol{\theta})\frac{\partial S(L_i\mid Z_i,\boldsymbol{\theta})/\partial \beta-\partial S(R_i\mid Z_i,\boldsymbol{\theta})/\partial \beta}{S( L_i\mid Z_i,\boldsymbol{\theta})-S(R_i\mid Z_i,\boldsymbol{\theta})}\cdot
\end{align*}
Then, we can show that

\begin{align*}
\frac{\partial }{\partial a_k}\sum_{l=k+1}^K C_{i,l}(\boldsymbol{\theta})&=\frac{(c_k\vee L_i-c_{k-1})e^{a_{i,k}}\sum_{l=k}^KS(c_l\vee L_i\mid Z_i,\boldsymbol{\theta})}{S( L_i\mid Z_i,\boldsymbol{\theta})-S(R_i\mid Z_i,\boldsymbol{\theta})}\\
&\quad -\frac{(c_k\wedge R_i-c_{k-1})e^{a_{i,k}}I(R_i\geq c_{k-1})\sum_{l=k+1}^KS(c_l\vee R_i\mid Z_i,\boldsymbol{\theta})}{S( L_i\mid Z_i,\boldsymbol{\theta})-S(R_i\mid Z_i,\boldsymbol{\theta})}
\end{align*}
\begin{align*}
&\quad-\sum_{l=k+1}^K C_{i,l}(\boldsymbol{\theta})\frac{\partial S(L_i\mid Z_i,\boldsymbol{\theta})/\partial a_k-\partial S(R_i\mid Z_i,\boldsymbol{\theta})/\partial a_k}{S( L_i\mid Z_i,\boldsymbol{\theta})-S(R_i\mid Z_i,\boldsymbol{\theta})}\cdot
\end{align*}
We now introduce:
\begin{align*}
E_{i,k}&=\exp(-a_{i,k}-e^{a_{i,k}}c_{k-1}\!\vee L_i)+\big(\exp(-a_{i,k})+c_{k-1}\!\vee L_i\big)\big(\exp(a_{i,k}-e^{a_{i,k}}c_{k-1}\!\vee L_i)c_{k-1}\!\vee L_i\big)\\
&\quad + \exp(-a_{i,k}-e^{a_{i,k}}c_{k-1}\vee L_i)+\big(\exp(-a_{i,k})+c_{k}\wedge R_i\big)\big(\exp(a_{i,k}-e^{a_{i,k}}c_{k}\wedge R_i)c_{k}\vee R_i\big),
\end{align*}
such that
\begin{align*}
\frac{\partial D_{i,k}(\boldsymbol{\theta})}{\partial a_k}&=- \frac{E_{i,k}\exp\big(e^{a_{i,k}}c_{k-1}-\sum_{j=1}^{k-1}e^{a_{i,j}}(c_j-c_{j-1})\big)J_{k,i}}{S(L_i\mid Z_i,\boldsymbol\theta)-S(R_i\mid Z_i,\boldsymbol\theta)}+D_{i,k}(\boldsymbol{\theta})e^{a_{i,k}}c_{k-1}J_{k,i}\\
&\quad -D_{i,k}(\boldsymbol{\theta})\frac{\partial S(L_i\mid Z_i,\boldsymbol{\theta})/\partial a_k-\partial S(R_i\mid Z_i,\boldsymbol{\theta})/\partial a_k}{S( L_i\mid Z_i,\boldsymbol{\theta})-S(R_i\mid Z_i,\boldsymbol{\theta})}J_{k,i},\\
\frac{\partial D_{i,k}(\boldsymbol{\theta})}{\partial \beta}&=-Z_i\frac{E_{i,k}\exp\big(e^{a_{i,k}}c_{k-1}-\sum_{j=1}^{k-1}e^{a_{i,j}}(c_j-c_{j-1})\big)J_{k,i}}{S(L_i\mid Z_i,\boldsymbol\theta)-S(R_i\mid Z_i,\boldsymbol\theta)}\\
&\quad +Z_iD_{i,k}(\boldsymbol{\theta})(e^{a_{i,k}}c_{k-1}-\sum_{j=1}^{k-1}e^{a_{i,j}}(c_j-c_{j-1}))J_{k,i}\\
&\quad -D_{i,k}(\boldsymbol{\theta})J_{k,i}\frac{\partial S(L_i\mid Z_i,\boldsymbol{\theta})/\partial \beta-\partial S(R_i\mid Z_i,\boldsymbol{\theta})/\partial \beta}{S( L_i\mid Z_i,\boldsymbol{\theta})-S(R_i\mid Z_i,\boldsymbol{\theta})}\cdot
\end{align*}
Finally, we have
\begin{align*}
\frac{\partial^2\mathrm{L}_1^{\mathrm{obs}}(\boldsymbol{\theta})}{\partial a_k^2}&=\frac{\partial C_{i,k}(\boldsymbol{\theta})}{\partial a_k}-e^{a_{i,k}}\Big(D_{i,k}(\boldsymbol{\theta})-c_{k-1}C_{i,k}(\boldsymbol{\theta})+\frac{\partial D_{i,k}(\boldsymbol{\theta})}{\partial a_k}-c_{k-1}\frac{\partial C_{i,k}(\boldsymbol{\theta})}{\partial a_k}\Big)\\
&\quad -e^{a_{i,k}}(c_k-c_{k-1})\left(\sum_{l=k+1}^K C_{i,l}(\boldsymbol{\theta})+\frac{\partial }{\partial a_k}\sum_{l=k+1}^K C_{i,l}(\boldsymbol{\theta})\right),\\
\frac{\partial^2\mathrm{L}_1^{\mathrm{obs}}(\boldsymbol{\theta})}{\partial a_k\partial \beta}&=Z_i\bigg\{\frac{\partial C_{i,k}(\boldsymbol{\theta})}{\partial a_k}-\frac{\partial C_{i,k}(\boldsymbol{\theta})}{\partial a_k}\sum_{j=1}^{k-1}e^{a_{i,j}}(c_j-c_{j-1})\\
&\quad-e^{a_{i,k}}\Big(D_{i,k}(\boldsymbol{\theta})-c_{k-1}C_{i,k}(\boldsymbol{\theta})+\frac{\partial D_{i,k}(\boldsymbol{\theta})}{\partial a_k}-c_{k-1}\frac{\partial C_{i,k}(\boldsymbol{\theta})}{\partial a_k}\Big)\bigg\},\\
\frac{\partial^2\mathrm{L}_1^{\mathrm{obs}}(\boldsymbol{\theta})}{\partial \beta^2}&=Z_i\bigg\{\frac{\partial C_{i,k}(\boldsymbol{\theta})^t}{\partial \beta}-\frac{\partial C_{i,k}(\boldsymbol{\theta})^t}{\partial \beta}\sum_{j=1}^{k-1}e^{a_{i,j}}(c_j-c_{j-1})\\
&\quad-e^{a_{i,k}}\Big(Z_i^tD_{i,k}(\boldsymbol{\theta})-c_{k-1}Z_i^tC_{i,k}(\boldsymbol{\theta})+\frac{\partial D_{i,k}(\boldsymbol{\theta})^t}{\partial \beta}-c_{k-1}\frac{\partial C_{i,k}(\boldsymbol{\theta})^t}{\partial \beta}\Big)\bigg\},
\end{align*}
and for the exact observations
\begin{align*}
\frac{\partial^2\mathrm{L}_2^{\mathrm{obs}}(\boldsymbol{\theta})}{\partial a_k^2}&=-\exp(a_{k}+\beta Z_i)R_{i,k},\\
\frac{\partial^2\mathrm{L}_2^{\mathrm{obs}}(\boldsymbol{\theta})}{\partial a_k\partial \beta}&=-Z_i\exp(a_{k}+\beta Z_i)R_{i,k},\\
\frac{\partial^2\mathrm{L}_2^{\mathrm{obs}}(\boldsymbol{\theta})}{\partial \beta^2}&=-Z_iZ_i^t\sum_{l=1}^K\bigg\{\exp(a_{l}+\beta Z_i)R_{i,l}\bigg\}.
\end{align*}


\begin{table}[!htb]
	\caption{\small Simulation results for the estimation of $\beta$ and $S_0$ in Scenarios S$3$ and S$4$. S$3$: $80\%$ of susceptible individuals. S$4$: $58\%$ of susceptible individuals. Among the susceptible individuals, $18\%$ of exact data, $19\%$ of left-censoring, $40\%$ of interval-censoring, $23\%$ of right-censoring.}\label{tab:simures3sup}
	\npdecimalsign{.}
	\nprounddigits{3}
	\small
	\begin{tabular}{| l| c| ccccccccc| }
		\hline 
		&& \multicolumn{9}{c| }{Adaptive Ridge estimate} \\  
		&$n$& \footnotesize Bias($\hat \beta$) & \footnotesize SE($\hat \beta$) &\footnotesize MSE($\hat \beta$)& \footnotesize Bias($\hat \gamma$) & \footnotesize SE($\hat \gamma$)&\footnotesize MSE($\hat \gamma$)&\footnotesize $\mathrm{IBias}^2(\hat S_0)$& \footnotesize $\mathrm{IVar}(\hat S_0)$ & \footnotesize $\mathrm{TV}(\hat\lambda_0)$ \\ 
		\hline 
		S$3$&$200$ & \numprint{-0.01495} & \numprint{0.29074}& \numprint{0.08475325} & \numprint{0.10182047} &\numprint{0.4983184}&\numprint{0.2586886} & \numprint{0.00356} & \numprint{0.32439}&  \numprint{0.83967}\\
		&&\numprint{0.00318} &\numprint{0.23611} &\numprint{0.05575804}&\numprint{0.01148162} &\numprint{0.6304594}&\numprint{0.3976109} & & & \\ 
		\hline 
		&$400$ &  \numprint{-0.01666} &\numprint{0.20734}&\numprint{0.04326743} &\numprint{0.07463} &\numprint{0.35572}&\numprint{0.1321064} &\numprint{0.00183}  & \numprint{0.15971}& \numprint{0.65867}\\
		&& \numprint{-0.00471} &\numprint{0.16212}&\numprint{0.026305082} &\numprint{0.02730} &\numprint{0.43336}&\numprint{0.1885462} & & & \\ 
		\hline 
		&$1\,000$ & \numprint{0.00583}  & \numprint{0.12701}& \numprint{0.01616553} & \numprint{0.02518} & \numprint{0.18435}& \numprint{0.03461895} & \numprint{0.00052} &\numprint{0.05948} & \numprint{0.41392}\\
		&& \numprint{0.00553} & \numprint{0.09446}& \numprint{0.008953273} & \numprint{0.01150} & \numprint{0.19751}& \numprint{0.03914245}& & & \\ 
		\hline
		\hline 
		S$4$&$200$ & \numprint{-0.02066} & \numprint{0.38697}& \numprint{0.1501726} & \numprint{0.07654} &\numprint{0.47909}&\numprint{0.2353856} & \numprint{0.00469} & \numprint{0.56310}&  \numprint{1.19482}\\
		&&\numprint{-0.00974} &\numprint{0.30966}&\numprint{0.09598418} &\numprint{0.03812} &\numprint{0.51093}&\numprint{0.2624719} & & & \\ 
		\hline 
		&$400$ &  \numprint{-0.02277} &\numprint{0.25531}&\numprint{0.06570167} &\numprint{0.04775} &\numprint{0.29615}&\numprint{0.08998489} &\numprint{0.00287}  & \numprint{0.25528}& \numprint{0.81049}\\
		&& \numprint{0.00285} &\numprint{0.20912}&\numprint{0.0437393} &\numprint{0.01621} &\numprint{0.30913}&\numprint{0.09582412} & & & \\ 
		\hline 
		&$1\,000$ & \numprint{-0.00860}  & \numprint{0.14980}& \numprint{0.022514} & \numprint{0.03174} & \numprint{0.18597}& \numprint{0.03559227} & \numprint{0.00086} &\numprint{0.09605} & \numprint{0.53034}\\
		&& \numprint{0.00791} & \numprint{0.12385}& \numprint{0.01540139} & \numprint{0.00437} & \numprint{0.20537}& \numprint{0.04219593}& & & \\ 
		\hline 
	\end{tabular}
\end{table}

\end{document}